\documentclass[
  reprint,
  superscriptaddress,
  amsmath,
  amssymb,
  aps,
  pra,
  floatfix,
]{revtex4-2}

\usepackage{graphicx}
\usepackage{dcolumn}
\usepackage{bm}
\usepackage{braket}
\usepackage[
  colorlinks=true,
  urlcolor=blue,
  citecolor=blue,
  linkcolor=blue
]{hyperref}
\usepackage[dvipsnames]{xcolor}
\usepackage{mathtools}
\usepackage{mathrsfs}
\usepackage{array}
\usepackage{booktabs}
\usepackage{enumerate}
\usepackage{siunitx}
\usepackage{subcaption}
\usepackage{caption}
\usepackage{ragged2e}

\DeclareCaptionJustification{justifiedhyphen}{\justifying}
\begin{document}

\title{Efficient calculation of the full Laguerre-Gaussian spectrum of entangled biphotons}

\author{Jiabao Peng}
\email{jiabao.peng@uni-jena.de}
\affiliation{%
  Theoretisch-Physikalisches Institut, Friedrich-Schiller-Universität Jena, Max-Wien-Platz 1,
  D-07743 Jena, Germany}
\affiliation{
  Helmholtz-Institut Jena, Fröbelstieg 3, D-07743 Jena, Germany
}

\author{Richard Bernecker}
\affiliation{%
  Theoretisch-Physikalisches Institut, Friedrich-Schiller-Universität Jena, Max-Wien-Platz 1,
  D-07743 Jena, Germany}
\affiliation{
  Helmholtz-Institut Jena, Fröbelstieg 3, D-07743 Jena, Germany
}

\author{Stephan Fritzsche}
\affiliation{%
  Theoretisch-Physikalisches Institut, Friedrich-Schiller-Universität Jena, Max-Wien-Platz 1,
  D-07743 Jena, Germany}
\affiliation{
  Helmholtz-Institut Jena, Fröbelstieg 3, D-07743 Jena, Germany
}
\affiliation{
  GSI Helmholtzzentrum für Schwerionenforschung GmbH, Planckstrasse 1, D-64291 Darmstadt,
  Germany
}

\begin{abstract}
  The spatial biphoton state generated by spontaneous parametric down-conversion (SPDC)
  enables applications in high-dimensional quantum information processing and is often
  described in Laguerre–Gaussian (LG) modes.
  However, calculating the full LG spectrum for arbitrary radial and orbital angular momentum
  (OAM) indices requires computationally demanding multidimensional integrations.
  We present a generating-function approach that analytically performs the four-dimensional
  transverse spatial integral without approximating the phase-matching function.
  The source geometry can be completely described by a few dimensionless parameters, which
  reduce to two parameters for wave-vector-degenerate photons with equal waists.
  Our derived expressions for the full LG spectrum allow for computationally efficient and
  numerically stable evaluation at high mode orders, addressing a crucial obstacle in previous
  works.
  To illustrate our approach, we compare OAM spectra obtained by projection onto the lowest
  radial order and by radially insensitive detection, and investigate how high-dimensional
  maximal entanglement can be approached by tuning source parameters.
  Our results provide a new analytical tool for efficiently characterizing and engineering
  high-dimensional spatial entanglement in SPDC.
\end{abstract}

\maketitle

\section{Introduction}
High-dimensional entangled states in degrees of freedom (DoFs) with dimensionality $d>2$ are
important resources for high-capacity quantum communication~\cite{qm}, sensitive quantum
imaging~\cite{PhysRevLett.92.233601,PhysRevLett.92.033601,PhysRevA.52.R3429}, and efficient
quantum computing~\cite{abe8770,Chen_2021,RevModPhys.95.045006}.
Spontaneous parametric down-conversion (SPDC), where a pump photon converts into a pair of
signal and idler photons, is widely used to generate biphoton states entangled in photon
number~\cite{Bimbard2010,Thekkadath2020},
polarization~\cite{PhysRevLett.93.230503,Fedorov_2011,Sekga2023}, spatial
modes~\cite{10.1098/0442rsta.2015.,10.1116/1.5112027}, temporal and frequency
modes~\cite{PhysRevLett.118.110501,Kues2017,Serino:24}, and time
bins~\cite{PhysRevLett.111.153602,Grassani:15}.
Among these DoFs, orbital angular momentum (OAM) is particularly attractive for
high-dimensional entanglement.
The large number of experimentally accessible OAM modes~\cite{Dada2011HighDimOAM}, together
with recent advances in their generation~\cite{Mair2001,Kovlakov2018} and
detection~\cite{Mirhosseini2013OAMSorter,Karan2025BroadbandOAM}, makes OAM a prime candidate
for realizing high-dimensional photonic entanglement~\cite{Erhard2020,Zhang:24}.

The spatial biphoton state can be conveniently represented in the Laguerre-Gaussian (LG)
basis~\cite{Allen1992OAMLG,Torres2003SpiralBandwidth}.
The azimuthal index \(\ell\) specifies the OAM carried by the LG mode, and the radial index
\(p\) describes its radial profile.
Experimentally, the LG spectrum is the distribution of coincidence probabilities across
pairs of signal and idler LG modes, inferred from the coincidence counts recorded for each
mode pair~\cite{SevillaGutierrez2024}.
Each probability is obtained by the corresponding overlap amplitude between the biphoton state and
the joint signal–idler LG modes.
Calculating these amplitudes requires a four-dimensional transverse integral determined by
the pump-beam profile and the phase-matching function (PMF), making its evaluation
challenging.

Analytical treatments of LG spectra often simplify the PMF using Gaussian or super-Gaussian
approximations~\cite{PhysRevLett.92.127903,Miatto2012}, or employ the thin-crystal limit,
where the PMF is effectively unity~\cite{PhysRevA.103.063508}.
While these approaches considerably simplify the LG mode projection, they neglect the
oscillatory structure of the full PMF, potentially reducing accuracy in thick crystals,
under strong focusing, and for higher-order LG modes.
Coincidence amplitudes for arbitrary LG pump modes have been derived~\cite{Yao_2011} in the
thin-crystal limit.
The validity of this approximation can be quantified by a dimensionless focusing
parameter~\cite{PhysRevA.103.063508}.
Dimensionless parameters have also been introduced to characterize SPDC for Gaussian-mode
coupling ~\cite{Bennink2010}, source brightness optimization ~\cite{Coccia2023Focusing}, or
OAM modes that are restricted to the lowest radial order ~\cite{Miatto2012OAMBandwidth}.
These treatments, however, cannot describe the full LG spectrum beyond the thin-crystal
regime.

More general treatments can accommodate arbitrary pump beams and describe spatial-spectral
mode coupling~\cite{Osorio2008Spatiotemporal,PhysRevA.106.063711,Bernecker2023FocalPlanes},
but the resulting expressions involve nested finite sums and special functions.
Their direct evaluation can become numerically unstable at high mode orders.
Alternatively, the completeness relation for radial modes yields an exact integral
representation~\cite{PhysRevA.97.063846}.
This simplifies the calculation but does not individually resolve amplitudes for every
radial order.
Engineering high-dimensional quantum states necessitates an efficient and numerically stable
description of the full LG spectrum~\cite{Kovlakov2018,Valencia2021}.

Here we develop a generating-function formulation that evaluates the transverse integrals in
the LG coincidence amplitudes analytically while keeping the full phase-matching function.
The remaining calculation involves coefficient extraction and a one-dimensional longitudinal
integral.
In this way, amplitudes for arbitrary pump, signal, and idler LG indices can be obtained
within a compact description in terms of dimensionless parameters.
In Sec.~II, we outline the mathematical formulation of our method, identify the
dimensionless parameters describing the source geometry, and assess its computational
performance.
In Sec.~III, we examine how the source parameters and the treatment of radial modes affect
the OAM spectra and fidelities with maximally entangled states (MES) as targets.
We also investigate the high-order radial-mode distributions for different pump modes.

\section{Theoretical model}
\subsection{Biphoton state from SPDC}

We consider SPDC in a second-order nonlinear crystal with susceptibility $\chi^{(2)}$.
We treat the pump as a classical, undepleted field~\cite{KellerRubin1997,WALBORN201087}, and
we work in the low-gain regime, where we neglect the generation of multiple
pairs~\cite{KellerRubin1997,WALBORN201087}.
The pump beam propagates along the $z$ axis and is focused at the center of the nonlinear
crystal.

In the quasi-collinear and paraxial regime~\cite{Monken1998AngularSpectrum,WALBORN201087},
the wave vector of each field can be decomposed as
\[
  \mathbf{k}_j=\mathbf{q}_j+k_{j,z}\mathbf{e}_z,\qquad j\in\{\mathrm p,\mathrm s,\mathrm i\},
\]
where $\mathbf{q}_j$ is the transverse wave vector and $k_{j,z}$ is the longitudinal
wave-vector component.
We assume narrowband detection centered at the frequencies $\omega_{j0}$, so that
$k_j=k_j(\omega_{j0})$ is fixed for each field.
Under the paraxial approximation $|\mathbf{q}_j|\ll k_j$ holds, such that the longitudinal
component can be written as
\begin{equation*}
  k_{j,z}(\mathbf{q}_j) \simeq k_j - \frac{|\mathbf{q}_j|^2}{2k_j}.
\end{equation*}
The central frequencies satisfy energy conservation
$\omega_{\mathrm p}=\omega_{\mathrm s}+\omega_{\mathrm i}$~\cite{KellerRubin1997,WALBORN201087}.
Moreover, when the transverse size of the crystal is much larger than the pump waist,
transverse wave-vector conservation gives
$\mathbf{q}_{\mathrm p}=\mathbf{q}_{\mathrm s}+\mathbf{q}_{\mathrm i}$~\cite{Monken1998AngularSpectrum,WALBORN201087}.
With these assumptions, the biphoton state in momentum-space representation
\cite{PhysRevA.62.043816,WALBORN201087,Karan2020} can be written as
\begin{equation}
  \begin{aligned}
    \ket{\psi} =&\mathcal{N} \int d^2\mathbf{q}_{\mathrm s}\,d^2\mathbf{q}_{\mathrm i}\,
    V_{\mathrm p}(\mathbf{q}_{\mathrm s}+\mathbf{q}_{\mathrm i})\,
    \Phi_{\mathrm{PM}}(\Delta k_z)\, \\
    &\times \hat{a}_{\mathrm s}^\dagger(\mathbf{q}_{\mathrm s}) \,
    \hat{a}_{\mathrm i}^\dagger(\mathbf{q}_{\mathrm i}) \ket{\mathrm{vac}},
  \end{aligned}
  \label{eq:biphoton_state_q}
\end{equation}
The function $V_{\mathrm p}(\mathbf{q}_{\mathrm s}+\mathbf{q}_{\mathrm i})$ is the angular
spectrum of the pump beam, describing its spatial distribution.
$\Phi_{\mathrm{PM}}(\Delta k_z)$ is the PMF of the crystal.
Equation~\eqref{eq:biphoton_state_q} represents the normalized two-photon component
conditioned on pair generation.
The vacuum component and the absolute pair-generation amplitude of low-gain SPDC are
omitted.
The global normalization constant $\mathcal{N}$ can be evaluated directly in momentum space
as
\begin{equation*}
  |\mathcal N|^{-2} = \int d^2\mathbf q_{\mathrm s}\,d^2\mathbf q_{\mathrm i}\, \left|
  V_{\mathrm p}(\mathbf q_{\mathrm s}+\mathbf q_{\mathrm i}) \Phi_{\mathrm{PM}}(\Delta k_z)
  \right|^2.
\end{equation*}

Angle tuning in birefringent bulk crystals can achieve collinear phase
matching~\cite{Karan2020,Nape2023QuantumStructuredLight}.
The total longitudinal mismatch is
$k_{\mathrm p,z}-k_{\mathrm s,z}-k_{\mathrm i,z}=\Delta k_0+\Delta k_z$, where
$\Delta k_0=k_{\mathrm p}-k_{\mathrm s}-k_{\mathrm i}$ is the constant central mismatch and
$\Delta k_z$ contains the contributions from the transverse wave vectors.
We absorb the phase associated with $\Delta k_0$ into the effective profile
$\chi_{\mathrm{eff}}^{(2)}(z)=\chi^{(2)}(z)e^{i\Delta k_0z}$.
For periodically poled crystals, this profile includes both the domain-inversion pattern and
the central-mismatch phase.
For first-order quasi-phase matching, the grating harmonic with wave vector $-2\pi/\Lambda$
compensates the central mismatch, giving $\Delta k_0=2\pi/\Lambda$, where $\Lambda$ is the
poling period~\cite{Baghdasaryan2023DomainEngineering}.
A constant effective profile for such a crystal additionally assumes that only the
phase-matched grating harmonic is included.

The remaining mismatch $\Delta k_z$ is therefore the paraxial contribution from the
transverse wave vectors,
\begin{equation}
  \begin{aligned}
    \Delta k_z ={}&-\frac{|\mathbf{q}_{\mathrm s}+\mathbf{q}_{\mathrm i}|^2}{2k_{\mathrm p}}
    +\frac{|\mathbf{q}_{\mathrm s}|^2}{2k_{\mathrm s}}
    +\frac{|\mathbf{q}_{\mathrm i}|^2}{2k_{\mathrm i}}.
  \end{aligned}
  \label{eq:delta_kz}
\end{equation}
Phase-matching (PM) configurations are classified as types 0, I, and II according to the
polarizations of the pump, signal, and idler
photons~\cite{Nape2023QuantumStructuredLight,Schneeloch2019SPDC}.
With the central-mismatch phase already included in $\chi_{\mathrm{eff}}^{(2)}(z)$, the PM function
is its Fourier transform with respect to the remaining mismatch $\Delta k_z$:
\begin{equation}
  \Phi_{\mathrm{PM}}(\Delta k_z) = \int_{-L/2}^{L/2} dz\, \chi_{\mathrm{eff}}^{(2)}(z)
  e^{i\Delta k_z z},
  \label{eq:pmf}
\end{equation}
where $L$ is the length of the nonlinear crystal.
Engineering the longitudinal susceptibility profile $\chi^{(2)}(z)$ allows the
phase-matching function and the resulting biphoton state to be
tailored~\cite{Baghdasaryan2023DomainEngineering,Bernecker2024MaximallyEntangled}.

\subsection{Expansion in Laguerre--Gaussian basis}

We take the pump angular spectrum to be a single LG mode,
\begin{equation*}
  V_{\mathrm p}(\mathbf{q}_{\mathrm p}) =
  \mathrm{LG}^{\ell_{\mathrm p}}_{p_{\mathrm p}}(\mathbf{q}_{\mathrm p},w_{\mathrm p}),
\end{equation*}
where $p_{\mathrm p} \in \mathbb{N}_0$ is the radial index,
$\ell_{\mathrm p} \in \mathbb{Z}$ is the OAM index, and $w_{\mathrm p}$ is the pump waist.
Each LG basis mode is an OAM eigenmode with an angular-momentum projection of $\ell\hbar$
per photon along the propagation axis.
Modes with $\ell\ne0$ have an azimuthally twisted phase distribution.
The radial index $p$ specifies the number of radial nodes in the transverse plane.

The momentum-space representation of a Laguerre--Gaussian mode at $z=0$, the crystal center,
is
\begin{equation}
  \begin{aligned}
    &\mathrm{LG}^{\ell}_{p}(\mathbf{q},w) =\sqrt{\frac{w^2 p!}{2\pi(p+|\ell|)!}}
    (-1)^p\left(\frac{|\mathbf{q}|w}{\sqrt{2}}\right)^{|\ell|} \\
    &\times \exp\left(-\frac{|\mathbf{q}|^2w^2}{4}\right)L_p^{|\ell|}
    \left(\frac{|\mathbf{q}|^2w^2}{2}\right)e^{i\ell \arg(\mathbf{q})},
  \end{aligned}
  \label{eq:lg_momentum}
\end{equation}
where $L_p^{|\ell|}(\cdot)$ denotes the generalized Laguerre polynomial.

The biphoton state \eqref{eq:biphoton_state_q} can then be expanded in the LG basis as
\begin{equation}
  \ket{\psi} = \sum_{p_{\mathrm s},\ell_{\mathrm s}} \sum_{p_{\mathrm i},\ell_{\mathrm i}}
  C_{p_{\mathrm p},p_{\mathrm s},p_{\mathrm i}}^{\ell_{\mathrm p},\ell_{\mathrm s},\ell_{\mathrm i}}
  \, \ket{p_{\mathrm s},\ell_{\mathrm s}} \ket{p_{\mathrm i},\ell_{\mathrm i}}.
  \label{eq:biphoton_lg_expansion}
\end{equation}
Here
$\ket{p,\ell}=\int d^2\mathbf{q}\,\mathrm{LG}^{\ell}_{p}(\mathbf{q},w)\hat{a}^{\dagger}(\mathbf{q})\ket{\mathrm{vac}}$
is the single-photon representation of the LG mode $\mathrm{LG}^{\ell}_{p}(\mathbf{q},w)$.
The LG-basis expansion coefficients, also called coincidence amplitudes, are obtained by
projecting the biphoton state onto the signal and idler LG modes:
\begin{equation}
  \begin{aligned}
    C_{p_{\mathrm p},p_{\mathrm s},p_{\mathrm i}}^{\ell_{\mathrm p},\ell_{\mathrm s},\ell_{\mathrm i}}
    &= \Bigl( \bra{p_{\mathrm s},\ell_{\mathrm s}} \bra{p_{\mathrm i},\ell_{\mathrm i}} \Bigl)\ket{\psi} \\
    &= \mathcal{N}\int d^2\mathbf{q}_{\mathrm s}\,d^2\mathbf{q}_{\mathrm i}\,\mathrm{LG}^{\ell_{\mathrm p}}_{p_{\mathrm p}}
    (\mathbf{q}_{\mathrm s}+\mathbf{q}_{\mathrm i},w_{\mathrm p})\,\Phi_{\mathrm{PM}}(\Delta k_z) \\
    &\times\left[\mathrm{LG}^{\ell_{\mathrm s}}_{p_{\mathrm s}}(\mathbf{q}_{\mathrm s},w_{\mathrm s})\right]^*
    \left[\mathrm{LG}^{\ell_{\mathrm i}}_{p_{\mathrm i}}(\mathbf{q}_{\mathrm i},w_{\mathrm i})\right]^*.
  \end{aligned}
  \label{eq:lg_projection_coefficient}
\end{equation}
The probability of detecting the signal and idler photons in the LG modes
\((p_{\mathrm s},\ell_{\mathrm s})\) and \((p_{\mathrm i},\ell_{\mathrm i})\), respectively,
is therefore
\begin{equation*}
  P_{p_{\mathrm s},p_{\mathrm i}}^{\ell_{\mathrm s},\ell_{\mathrm i}}(p_{\mathrm p},\ell_{\mathrm p}) =
  \left|C_{p_{\mathrm p},p_{\mathrm s},p_{\mathrm i}}^{\ell_{\mathrm p},\ell_{\mathrm s},\ell_{\mathrm i}}\right|^2.
\end{equation*}
For fixed pump indices, the complete set of probabilities
\(P_{p_{\mathrm s},p_{\mathrm i}}^{\ell_{\mathrm s},\ell_{\mathrm i}}(p_{\mathrm p},\ell_{\mathrm p})\)
is referred to as the LG spectrum.

A radially insensitive OAM measurement resolves the signal and idler OAM indices
\((\ell_{\mathrm s},\ell_{\mathrm i})\) while tracing over their radial LG indices
\((p_{\mathrm s},p_{\mathrm i})\) \cite{Kulkarni2017SingleShot}.
The probability associated with a given pair of signal and idler OAM indices therefore
receives incoherent contributions from all radial-mode pairs.
For a pump prepared in the mode \((p_{\mathrm p},\ell_{\mathrm p})\), the resulting radially
insensitive OAM spectrum is
\begin{equation}
  P_{\mathrm{tr}}^{\ell_{\mathrm s},\ell_{\mathrm i}}(p_{\mathrm p},\ell_{\mathrm p})
  =\sum_{p_{\mathrm s}=0}^{\infty}\sum_{p_{\mathrm i}=0}^{\infty}
  \left|C_{p_{\mathrm p},p_{\mathrm s},p_{\mathrm i}}^{\ell_{\mathrm p},\ell_{\mathrm s},\ell_{\mathrm i}}\right|^2 .
  \label{eq:radially_insensitive_oam_spectrum}
\end{equation}

The angular Schmidt spectrum studied in Ref.~\cite{PhysRevA.97.063846} is a radially
insensitive quantity of this form.
There, radial-mode completeness removes the explicit infinite sums, although the resulting
expression still involves a four-dimensional transverse integral.
More recently, a broadband OAM detector enabled measurement of the spectrum without
radial-mode postselection \cite{Karan2025BroadbandOAM}.
In our approach, the transverse Gaussian integration will be performed analytically,
allowing Eq.~\eqref{eq:radially_insensitive_oam_spectrum} to be evaluated efficiently.
The direct calculation is detailed in Appendix~\ref{app:direct-radial-trace}.

\subsection{Generating-function approach and reduction of the coincidence amplitudes}
In this section, we summarize the mathematical operations used to reduce the coincidence
amplitudes.
The transverse integral is evaluated first, after which the OAM and radial indices are
selected by coefficient extraction, leaving only a one-dimensional integral along the
longitudinal direction over the crystal length.
The OAM extraction also recovers the OAM conservation rule that is physically imposed by
rotational symmetry~\cite{Walborn2004OAMConservation}.
Full expressions and further intermediate steps are provided in
Appendices~\ref{app:gaussian-integral}--\ref{app:rho-expansion}.

Instead of evaluating Eq.~\eqref{eq:lg_projection_coefficient} separately for every
combination of LG indices by decomposing the LG modes into a sum representation like in
Ref.~\cite{Yao_2011}, we encode the complete LG basis in a single generating function.
Following Ref.~\cite{PhysRevA.83.053822}, we use
\begin{equation}
  \begin{aligned}
    G_\pm ={}&\frac{1}{1+\eta} \exp\Bigg[ \frac{i(q_x\pm i q_y)w\mu}{2(1+\eta)}
    -\frac{|\mathbf q|^2w^2(1-\eta)}{4(1+\eta)} \Bigg].
  \end{aligned}
  \label{eq:LG_generating_function_cartesian}
\end{equation}
Here \(\mathbf q=(q_x,q_y)\) is the transverse wave vector and \(w\) is the waist of the
corresponding LG mode.
In Eq.~\eqref{eq:LG_generating_function_cartesian}, the auxiliary variables \(\eta\) and
\(\mu\) generate the radial index and the magnitude of the OAM index, respectively, while
the signs $\pm$ select the OAM sign.
The mode with indices \((p,\ell)\) in Eq.~\eqref{eq:lg_momentum} is obtained by selecting
the corresponding coefficient,
\begin{equation}
  \mathrm{LG}^{\ell}_{p}(\mathbf{q},w) =N^{\ell}_{p}
  \left.\frac{1}{p!}\frac{\partial^p}{\partial\eta^p}
  \frac{\partial^{|\ell|}}{\partial\mu^{|\ell|}}
  G_{\operatorname{sgn}(\ell)}(\mathbf q,w;\eta,\mu) \right|_{\eta=\mu=0}
  \label{eq:LG_from_generating_function}
\end{equation}
with \(G_{\operatorname{sgn}(\ell)}=G_+\) for \(\ell>0\) and
\(G_{\operatorname{sgn}(\ell)}=G_-\) for \(\ell<0\).
For \(\ell=0\), the two choices are equivalent.
The normalization factor is
\(N^{\ell}_{p}= i^{-|\ell|}\bigl( w^2 2^{|\ell|}p!/(2\pi(p+|\ell|)!) \bigl)^{1/2}\).

To separate the longitudinal dependence from the transverse mode structure, we substitute
the phase-matching representation in Eq.~\eqref{eq:pmf} into the LG projection and introduce
the dimensionless longitudinal coordinate
\begin{equation}
  t=-\frac{2z}{L}, \qquad \widetilde{\chi}(t) = \frac{\chi_{\mathrm{eff}}^{(2)}(-Lt/2)}
  {\chi_0^{(2)}}.
  \label{eq:dimensionless_longitudinal_coordinate}
\end{equation}
The crystal is thereby mapped onto the interval $-1\leq t\leq1$, where $\chi_0^{(2)}$ is a
reference value of the second-order susceptibility, making $\widetilde{\chi}(t)$
dimensionless.
We use the shorthand
\(G_j(\mathbf q)\equiv G_{\operatorname{sgn}(\ell_j)}(\mathbf q,w_j;\eta_j,\mu_j)\), with
\(j\in\{\mathrm p,\mathrm s,\mathrm i\}\).
We denote the resulting expression by
\begin{equation}
  \begin{aligned}
    \mathscr{Z}(\boldsymbol\eta,\boldsymbol\mu) ={}& \int_{-1}^{1}dt\, \widetilde{\chi}(t)
    \int d^2\mathbf q_{\mathrm s}\,d^2\mathbf q_{\mathrm i} \\
    &\times G_{\mathrm p}(\mathbf q_{\mathrm s}+\mathbf q_{\mathrm i})
    G_{\mathrm s}^*(\mathbf q_{\mathrm s}) G_{\mathrm i}^*(\mathbf q_{\mathrm i})
    e^{-iL\Delta k_z t/2}.
  \end{aligned}
  \label{eq:generating_amplitude}
\end{equation}
The stars on $G_{\mathrm s}$ and $G_{\mathrm i}$ denote coefficientwise complex conjugation.

The physical LG coefficient is recovered using the coefficient extraction
operator~\cite{Withers2010MultivariateBell} as
\begin{equation*}
  C_{p_{\mathrm p},p_{\mathrm s},p_{\mathrm i}}^{\ell_{\mathrm p},\ell_{\mathrm s},\ell_{\mathrm i}} \propto
  [\eta_{\mathrm p}^{p_{\mathrm p}}\eta_{\mathrm s}^{p_{\mathrm s}}\eta_{\mathrm i}^{p_{\mathrm i}}]
  [\mu_{\mathrm p}^{|\ell_{\mathrm p}|}\mu_{\mathrm s}^{|\ell_{\mathrm s}|}\mu_{\mathrm i}^{|\ell_{\mathrm i}|}]
  \mathscr{Z}(\boldsymbol\eta,\boldsymbol\mu),
\end{equation*}
where $\boldsymbol\eta=(\eta_{\mathrm p},\eta_{\mathrm s},\eta_{\mathrm i})$ and
$\boldsymbol\mu=(\mu_{\mathrm p},\mu_{\mathrm s},\mu_{\mathrm i})$, and the normalization
factors are handled below.

\subsubsection{Transverse integration}
To evaluate the transverse integral, we introduce the variables
$\mathbf q_+=\mathbf q_{\mathrm s}+\mathbf q_{\mathrm i},\quad\mathbf q_-=\mathbf q_{\mathrm s}-\mathbf q_{\mathrm i}.$
Using Eq.~\eqref{eq:delta_kz}, the phase-mismatch is quadratic in the transverse wave
vectors.
After inserting the generating functions, the complete dependence on the transverse wave
vectors is therefore Gaussian and can be written in the form
{\small
  \begin{equation*}
    \begin{aligned}
      \exp\Big[ &-A_+(t;\boldsymbol\eta)|\mathbf q_+|^2 -A_-(t;\boldsymbol\eta)|\mathbf q_-|^2
      -A_{+-}(t;\boldsymbol\eta)\, \mathbf q_+\cdot\mathbf q_- \\
      &+\mathbf J_+ (\boldsymbol\eta,\boldsymbol\mu) \cdot\mathbf q_+ +\mathbf J_-
      (\boldsymbol\eta,\boldsymbol\mu) \cdot\mathbf q_-\Big].
    \end{aligned}
  \end{equation*}
}
With
\(\Delta(t;\boldsymbol\eta)=A_+(t;\boldsymbol\eta)A_-(t;\boldsymbol\eta)-A_{+-}^2(t;\boldsymbol\eta)/4\),
the four transverse integrations can be evaluated analytically.
We collect the numerator in
\begin{equation*}
  \begin{aligned}
    \mathcal Q_\perp(t;\boldsymbol\eta,\boldsymbol\mu)
    ={}&A_-(t;\boldsymbol\eta)\mathbf J_+^2(\boldsymbol\eta,\boldsymbol\mu)
    +A_+(t;\boldsymbol\eta)\mathbf J_-^2(\boldsymbol\eta,\boldsymbol\mu)\\
    &-A_{+-}(t;\boldsymbol\eta)\mathbf J_+(\boldsymbol\eta,\boldsymbol\mu)\cdot\mathbf J_-(\boldsymbol\eta,\boldsymbol\mu).
  \end{aligned}
\end{equation*}
The transverse integral then becomes
\begin{equation}
  \mathcal T(t;\boldsymbol\eta,\boldsymbol\mu)
  =\frac{\pi^2}{4\Delta(t;\boldsymbol\eta)} \times\frac{\exp\!\left[\dfrac{\mathcal Q_\perp(t;\boldsymbol\eta,\boldsymbol\mu)}{4\Delta(t;\boldsymbol\eta)}\right]}
  {(1+\eta_{\mathrm p})(1+\eta_{\mathrm s})(1+\eta_{\mathrm i})}.
  \label{eq:transverse_gaussian_result}
\end{equation}
The explicit forms of \(A_+(t;\boldsymbol\eta)\), \(A_-(t;\boldsymbol\eta)\),
\(A_{+-}(t;\boldsymbol\eta)\), \(\mathbf J_\pm(\boldsymbol\eta,\boldsymbol\mu)\), and
\(\Delta(t;\boldsymbol\eta)\) are given in Appendix~\ref{app:gaussian-integral}.

\subsubsection{OAM coefficient extraction}

To extract the OAM indices, we first rewrite the \(\boldsymbol\mu\)-dependent exponent in
Eq.~\eqref{eq:transverse_gaussian_result} in a form that can be expanded directly.
Collecting the coefficients of the three pairwise products of the azimuthal generating
variables, we write
\begin{equation*}
  \frac{\mathcal Q_\perp(t;\boldsymbol\eta,\boldsymbol\mu)}{4\Delta(t;\boldsymbol\eta)}
  =A\mu_{\mathrm p}\mu_{\mathrm s}+B\mu_{\mathrm p}\mu_{\mathrm i} +C\mu_{\mathrm s}\mu_{\mathrm i},
\end{equation*}
where the explicit expressions for \(A\), \(B\), and \(C\) are given in
Appendix~\ref{app:angular-extraction}.
The exponential can then be expanded as
\begin{equation}
  e^{A\mu_{\mathrm{p}}\mu_{\mathrm{s}}+B\mu_{\mathrm{p}}\mu_{\mathrm{i}}+C\mu_{\mathrm{s}}\mu_{\mathrm{i}}}
  =\sum_{a,b,c\geq0} \frac{A^aB^bC^c}{a!b!c!}
  \mu_{\mathrm{p}}^{a+b}\mu_{\mathrm{s}}^{a+c}\mu_{\mathrm{i}}^{b+c}.
  \label{eq:ABC_angular_expansion}
\end{equation}

The LG mode with index \(\ell_j\) is obtained by taking the coefficient of
\(\mu_j^{|\ell_j|}\), as prescribed by Eq.~\eqref{eq:LG_from_generating_function}.
Meanwhile, a term proportional to \(A^aB^bC^c\) in Eq.~\eqref{eq:ABC_angular_expansion}
carries the powers \(\mu_{\mathrm{p}}^{a+b}\mu_{\mathrm{s}}^{a+c}\mu_{\mathrm{i}}^{b+c}\).
Such a term can therefore survive the three coefficient extractions only if
\(a+b=|\ell_{\mathrm{p}}|\), \(a+c=|\ell_{\mathrm{s}}|\), and \(b+c=|\ell_{\mathrm{i}}|\).
Solving these three equations gives
\begin{equation}
  \makebox[0.8\columnwidth][c]{\resizebox{0.90\columnwidth}{!}{$\displaystyle
      a=\frac{|\ell_{\mathrm{p}}|+|\ell_{\mathrm{s}}|-|\ell_{\mathrm{i}}|}{2},\;
      b=\frac{|\ell_{\mathrm{p}}|+|\ell_{\mathrm{i}}|-|\ell_{\mathrm{s}}|}{2},\;
      c=\frac{|\ell_{\mathrm{s}}|+|\ell_{\mathrm{i}}|-|\ell_{\mathrm{p}}|}{2}.
  $}}
  \label{eq:angular_integers_ab}
\end{equation}

Let
\(\sigma_j=\operatorname{sgn}(\ell_j)\),  the Kronecker deltas contained in \(A\), \(B\), and \(C\) [Eq.~\eqref{app:ABC}] require
\(\sigma_{\mathrm{p}}=\sigma_{\mathrm{s}}\) whenever \(a>0\),
\(\sigma_{\mathrm{p}}=\sigma_{\mathrm{i}}\) whenever \(b>0\), and
\(\sigma_{\mathrm{s}}=-\sigma_{\mathrm{i}}\) whenever \(c>0\).
Consequently,
\begin{equation*}
  \begin{aligned}
    \ell_{\mathrm{s}}+\ell_{\mathrm{i}} &=\sigma_{\mathrm{s}}(a+c)+\sigma_{\mathrm{i}}(b+c)\\
    &=\sigma_{\mathrm{p}}a+\sigma_{\mathrm{p}}b +c(\sigma_{\mathrm{s}}+\sigma_{\mathrm{i}})
    =\sigma_{\mathrm{p}}(a+b) =\ell_{\mathrm{p}}.
  \end{aligned}
\end{equation*}
Thus, the angular coefficient extraction is nonzero only when
\(\ell_{\mathrm{p}}=\ell_{\mathrm{s}}+\ell_{\mathrm{i}}\).
In this formulation, OAM conservation follows directly from the powers and sign selection
rules of the generating function.

Once the OAM indices have been selected, the contribution for
\(p_{\mathrm{p}}=p_{\mathrm{s}}=p_{\mathrm{i}}=0\) can be collected in the kernel
\begin{equation}
  \resizebox{0.95\columnwidth}{!}{$\displaystyle
    \mathcal G(t) =
    \frac{(1+it\tan\phi_{\mathrm{i}})^a(1+it\tan\phi_{\mathrm{s}})^b(-1+it\tan\phi_{\mathrm{p}})^c}
    {[(1+it\tan\phi_1)(1+it\tan\phi_2)]^{a+b+c+1}}$}
  \label{eq:oam_gouy_kernel}
\end{equation}
The integers \(a\), \(b\), and \(c\) determine the powers with which the longitudinal
focusing parameters enter.
These parameters are defined by
\begin{equation}
  \tan\phi_j =\frac{L}{k_jw_j^2} =\frac{L}{2z_R^{(j)}},
  \qquad j\in\{\mathrm{p},\mathrm{s},\mathrm{i}\},
  \label{eq:focusing_phases}
\end{equation}
where \(z_R^{(j)}=k_jw_j^2/2\) is the Rayleigh distance of beam \(j\).
The two effective longitudinal focusing parameters \(\phi_1\) and \(\phi_2\) in the
denominator of Eq.~\eqref{eq:oam_gouy_kernel} arise from factorizing the determinant
\(\Delta(t;\mathbf 0)\) obtained from the transverse Gaussian integration.
Its dependence on \(t\) has the quadratic form \(\Delta(t;\mathbf 0)\propto D+iEt-Ft^2\),
which gives
\begin{equation}
  \tan\phi_{1,2} =\frac{E\pm\sqrt{E^2-4DF}}{2D}.
  \label{eq:effective_focusing_phases}
\end{equation}
The coefficients \(D\), \(E\), and \(F\), together with the factorization, are given in
Appendix~\ref{app:longitudinal-kernel}.

We can then describe the source geometry using the waist ratios
\begin{equation}
  \gamma_{\mathrm{s}}=\frac{w_{\mathrm{p}}}{w_{\mathrm{s}}}, \qquad
  \gamma_{\mathrm{i}}=\frac{w_{\mathrm{p}}}{w_{\mathrm{i}}},
  \label{eq:waist_ratios}
\end{equation}
together with the pump focusing parameter~\cite{Bennink2010} and the signal-to-idler
wave-number ratio,
\begin{equation}
  \xi_{\mathrm p}=\frac{L}{k_{\mathrm p}w_{\mathrm p}^2}, \qquad
  \nu=\frac{k_{\mathrm s}}{k_{\mathrm i}}.
  \label{eq:independent_focusing_nondegeneracy}
\end{equation}
Under wave-vector conservation, the longitudinal kernel \(\mathcal G(t)\) depends on the
source geometry only through these four independent dimensionless parameters,
\begin{equation}
  \boldsymbol\lambda_{\mathrm{geom}}
  =(\gamma_{\mathrm{s}},\gamma_{\mathrm{i}},\xi_{\mathrm p},\nu),
  \label{eq:geometric_controls}
\end{equation}
For fixed mode indices and a fixed normalized nonlinear profile, sources with the same
\(\boldsymbol\lambda_{\mathrm{geom}}\) therefore yield identical normalized coincidence
amplitudes.

\subsubsection{Radial coefficient expansion}
After transverse Gaussian integration and OAM coefficient extraction, we write the remaining
radial factor in exponential form,
\((\cdots)=e^{\ln(\cdots)}\equiv e^{\Psi(t;\boldsymbol\eta)}\).
The radial correction for the selected OAM indices is then
\begin{equation}
  \mathcal K_{\mathbf p}(t) =
  [\eta_{\mathrm p}^{p_{\mathrm p}}\eta_{\mathrm s}^{p_{\mathrm s}}\eta_{\mathrm i}^{p_{\mathrm i}}]
  \exp[\Psi(t;\boldsymbol\eta)-\Psi(t;\mathbf0)].
  \label{eq:radial_coefficient_extraction}
\end{equation}
The explicit form of \(\Psi(t;\boldsymbol\eta)\), including the radial correction to the
coupled denominator, is given in Eq.~\eqref{Psi} in Appendix~\ref{app:radial-extraction}.
For $\mathbf p=(p_{\mathrm p},p_{\mathrm s},p_{\mathrm i})=(0,0,0)$, the coefficient is
obtained by setting $\boldsymbol\eta=\mathbf0$.
Therefore,
\begin{equation*}
  \mathcal K_{\mathbf0}(t)=1.
\end{equation*}
To evaluate Eq.~\eqref{eq:radial_coefficient_extraction}, we define the partial derivatives
\begin{equation}
  \Psi_{\boldsymbol\alpha}(t) =\left. \partial_{\eta_{\mathrm p}}^{\alpha_{\mathrm p}}
  \partial_{\eta_{\mathrm s}}^{\alpha_{\mathrm s}}
  \partial_{\eta_{\mathrm i}}^{\alpha_{\mathrm i}} \Psi(t;\boldsymbol\eta)
  \right|_{\boldsymbol\eta=\mathbf0}, \qquad \boldsymbol\alpha\ne\mathbf0.
  \label{eq:radial_derivative_blocks}
\end{equation}
A derivative configuration \(\mathfrak r=\{n_{\boldsymbol\alpha}^{(\mathfrak r)}\}\)
specifies the multiplicities of partial derivatives whose multi-indices sum to the target
radial-index vector \(\mathbf p\).
The finite set \(\mathscr P(\mathbf p)\) contains all configurations satisfying
\begin{equation}
  \sum_{\boldsymbol\alpha\ne\mathbf0} n_{\boldsymbol\alpha}^{(\mathfrak r)}\boldsymbol\alpha
  =\mathbf p, \qquad n_{\boldsymbol\alpha}^{(\mathfrak r)}\in\mathbb N_0.
  \label{eq:radial_configuration_condition}
\end{equation}
With \(\boldsymbol\alpha!=\alpha_{\mathrm p}!\alpha_{\mathrm s}!\alpha_{\mathrm i}!\),
define
\begin{equation}
  \kappa_{\mathfrak r} =\prod_{\boldsymbol\alpha\ne\mathbf0}
  \frac{1}{n_{\boldsymbol\alpha}^{(\mathfrak r)}!
    (\boldsymbol\alpha!)^{n_{\boldsymbol\alpha}^{(\mathfrak r)}}}, \qquad R_{\mathfrak r}(t)
  =\prod_{\boldsymbol\alpha\ne\mathbf0}
  [\Psi_{\boldsymbol\alpha}(t)]^{n_{\boldsymbol\alpha}^{(\mathfrak r)}}.
  \label{eq:radial_configuration_factors}
\end{equation}
The complete radial correction is the finite sum
\begin{equation}
  \mathcal K_{\mathbf p}(t) =\sum_{\mathfrak r\in\mathscr P(\mathbf p)}
  \kappa_{\mathfrak r} R_{\mathfrak r}(t).
  \label{eq:finite_rho_expansion}
\end{equation}

Two examples make the role of the configurations explicit.
For \(\mathbf p=(0,1,1)\), a mixed partial derivative and a product of two first-order
partial derivatives contribute with unit weight:
\[
  \mathcal K_{0,1,1}(t) =\Psi_{(0,1,1)}(t) +\Psi_{(0,1,0)}(t)\Psi_{(0,0,1)}(t).
\]
For \(\mathbf p=(0,2,0)\), the second-order partial derivative and the squared first-order
partial derivative both carry a weight of \(1/2\):
\[
  \mathcal K_{0,2,0}(t) =\frac{1}{2}\Psi_{(0,2,0)}(t)
  +\frac{1}{2}\big[\Psi_{(0,1,0)}(t)\big]^2.
\]
The derivation is given in Appendix~\ref{app:rho-expansion}.

\subsubsection{Reduced coincidence amplitude}
The resulting coincidence amplitude is
\begin{equation}
  \begin{aligned}
    C_{p_{\mathrm p},p_{\mathrm s},p_{\mathrm i}}^{\ell_{\mathrm p},\ell_{\mathrm s},\ell_{\mathrm i}}
    &=\delta_{\ell_{\mathrm p},\ell_{\mathrm s}+\ell_{\mathrm i}}\,
    \mathcal A_{p_{\mathrm p},p_{\mathrm s},p_{\mathrm i}}^{\ell_{\mathrm p},\ell_{\mathrm s},\ell_{\mathrm i}}
    \int_{-1}^{1}dt\,\widetilde\chi(t)\mathcal G(t)\mathcal K_{\mathbf p}(t) \\
    &=\delta_{\ell_{\mathrm p},\ell_{\mathrm s}+\ell_{\mathrm i}}\,
    \mathcal A_{p_{\mathrm p},p_{\mathrm s},p_{\mathrm i}}^{\ell_{\mathrm p},\ell_{\mathrm s},\ell_{\mathrm i}}
    \sum_{\mathfrak r\in\mathscr P(\mathbf p)} \kappa_{\mathfrak r}\mathcal J_{\mathfrak r}.
  \end{aligned}
  \label{eq:final_normalized_amplitude}
\end{equation}
Here
\(\mathcal J_{\mathfrak r}=\int_{-1}^{1}dt\, \widetilde\chi(t)\mathcal G(t)R_{\mathfrak r}(t)\)
is the longitudinal contribution of each derivative configuration.
The second equality follows from Eq.~\eqref{eq:finite_rho_expansion}.
The Kronecker delta enforces OAM conservation arising from rotational symmetry, and
\(\mathcal A_{p_{\mathrm p},p_{\mathrm s},p_{\mathrm i}}^{\ell_{\mathrm p},\ell_{\mathrm s},\ell_{\mathrm i}}\)
is the normalization prefactor (Appendix~\ref{app:overall-prefactor}).
The function \(\widetilde\chi(t)\) describes the nonlinear longitudinal profile, while
\(\mathcal G(t)\) describes the longitudinal evolution for zero radial indices and
\(\mathcal K_{\mathbf p}(t)\) supplies the radial correction for the selected OAM indices.

\subsection{Computational efficiency}
Equation~\eqref{eq:final_normalized_amplitude} enables efficient calculation of the
coincidence amplitudes.
For high radial orders, we accelerate coefficient extraction using Cauchy sampling of the
generating function (Appendix~\ref{app:cauchy-extraction}).
We benchmark our method against a numerical implementation of the formulation by
Baghdasaryan \textit{et al.}~\cite{PhysRevA.106.063711} at several mode orders.
The wave-vector-degenerate and nondegenerate cases use the experimental parameters reported
by Sevilla-Guti\'errez \textit{et al.}~\cite{SevillaGutierrez2024} and Villar \textit{et
  al.}~\cite{Villar2019}, respectively.
\begin{center}
  \begin{minipage}{0.96\columnwidth}
    \captionof{table}{Runtime comparison.}
    \label{tab:main-runtime-comparison}
    \centering
    \small
    \resizebox{0.98\columnwidth}{!}{%
      \begin{tabular}{l c r r r}
        \toprule
        Configuration                          & $n=$ & Previous (s) & This work (s) & Speedup        \\
        \midrule
        Degenerate~\cite{SevillaGutierrez2024} & 2    & 0.0359       & 0.00116       & \(30.9\times\) \\
                                               & 3    & failed       & 0.00749       & --             \\
                                               & 5    & failed       & 0.127         & --             \\
                                               & 6    & failed       & 0.149         & --             \\
                                               & 9    & failed       & 0.462         & --             \\
        \midrule
        Nondegenerate~\cite{Villar2019}        & 2    & 0.161        & 0.0268        & \(6.00\times\) \\
                                               & 3    & failed       & 0.0660        & --             \\
                                               & 5    & failed       & 0.242         & --             \\
                                               & 6    & failed       & 0.322         & --             \\
                                               & 9    & failed       & 0.453         & --             \\
        \bottomrule
      \end{tabular}
    }
  \end{minipage}
\end{center}
In Table~\ref{tab:main-runtime-comparison}, order \(n\) corresponds to the mode indices
\((p_{\mathrm p},p_{\mathrm s},p_{\mathrm i}, \ell_{\mathrm p},\ell_{\mathrm s},\ell_{\mathrm i}) =(n-1,n,n,2n,3n+1,-n-1)\).
``Failed'' denotes non-finite results or excessive numerical errors, where
a dash indicates an unavailable speedup.
Our method is faster at low orders where both implementations give accurate results.
The computational cost of the previous formulation~\cite{PhysRevA.106.063711} grows rapidly
with mode order, and its evaluation may fail due to numerical instability in the
longitudinal integration of higher-order hypergeometric functions, whereas our method
remains tractable.
Further runtime and numerical-stability results are given in
Appendix~\ref{app:runtime-benchmark}, and scale-invariance validation in
Appendix~\ref{app:universal-collapse}.

\section{Results and Discussion}
\subsection{Comparison with previous analytical results}
In the wave-vector-degenerate case, \(k_{\mathrm s}=k_{\mathrm i}=k_{\mathrm p}/2\), we
assume a uniform longitudinal profile \(\widetilde\chi(t)=1\) while allowing different
signal and idler collection waists.
The effective focusing parameters then reduce to \(\phi_2=0\) and \(\phi_1\equiv\phi\), with
\[
  \tan\phi=\xi_{\mathrm p}
  \frac{4\gamma_{\mathrm s}^2\gamma_{\mathrm i}^2+\gamma_{\mathrm s}^2+\gamma_{\mathrm i}^2}
  {1+\gamma_{\mathrm s}^2+\gamma_{\mathrm i}^2}.
\]
For the lowest radial orders, \(p_{\mathrm p}=p_{\mathrm s}=p_{\mathrm i}=0\), no radial
derivatives are required, so \(\mathcal K_{\mathbf0}(t)=1\).
The resulting amplitudes allow a direct comparison with established expressions.
The ratios entering the longitudinal phase average are
\begin{equation*}
  \begin{aligned}
    \rho_{\mathrm{p}}= \frac{1+\gamma_{\mathrm{s}}^2+\gamma_{\mathrm{i}}^2}
    {4\gamma_{\mathrm{s}}^2\gamma_{\mathrm{i}}^2+\gamma_{\mathrm{s}}^2+\gamma_{\mathrm{i}}^2},\quad
    \rho_{\mathrm{s}}=2\gamma_{\mathrm{s}}^2\rho_{\mathrm{p}},\quad
    \rho_{\mathrm{i}}=2\gamma_{\mathrm{i}}^2\rho_{\mathrm{p}}.
  \end{aligned}
\end{equation*}
The full lowest-radial-order amplitude takes the form
\begin{equation}
  \resizebox{0.99\columnwidth}{!}{$
    \begin{aligned}
      &C^{\ell_{\mathrm{p}},\ell_{\mathrm{s}},\ell_{\mathrm{i}}}_{0,0,0}
      =\delta_{\ell_{\mathrm{p}},\ell_{\mathrm{s}}+\ell_{\mathrm{i}}} 2^{|\ell_{\mathrm{p}}|+c+2}
      \frac{2\phi}{\sqrt{\pi\tan\phi}}
      \frac{\sqrt{|\ell_{\mathrm{p}}|!\,|\ell_{\mathrm{s}}|!\,|\ell_{\mathrm{i}}|!}}
      {(|\ell_{\mathrm{s}}|-c)!\,(|\ell_{\mathrm{i}}|-c)!\,c!} \\
      &\quad\times
      \frac{\gamma_{\mathrm{s}}^{|\ell_{\mathrm{s}}|+1}\gamma_{\mathrm{i}}^{|\ell_{\mathrm{i}}|+1}}
      {(1+\gamma_{\mathrm{s}}^2+\gamma_{\mathrm{i}}^2)^{|\ell_{\mathrm{p}}|+c+1/2}
        \sqrt{4\gamma_{\mathrm{s}}^2\gamma_{\mathrm{i}}^2+\gamma_{\mathrm{s}}^2+\gamma_{\mathrm{i}}^2}}
      \frac{1}{2\phi}\int_{-\phi}^{\phi}d\theta \\
      &\quad\times
      [\rho_{\mathrm{i}}+(1-\rho_{\mathrm{i}})\cos\phi\,e^{i\theta}]^{|\ell_{\mathrm{s}}|-c}
      [\rho_{\mathrm{s}}+(1-\rho_{\mathrm{s}})\cos\phi\,e^{i\theta}]^{|\ell_{\mathrm{i}}|-c} \\
      &\quad\times [\rho_{\mathrm{p}}-(1+\rho_{\mathrm{p}})\cos\phi\,e^{i\theta}]^c.
    \end{aligned}
  $}
  \label{eq:p0_unequal_waists_phase_average}
\end{equation}
For \(\ell_{\mathrm p}>0\), Yao~\cite{Yao_2011} gives the following lowest-radial-order
coefficient in the notation of that work.
Here, \(\sigma_\ell=(|\ell_{\mathrm p}|+|\ell_{\mathrm s}|+|\ell_{\mathrm i}|)/2\).
\begin{equation*}
  \resizebox{0.99\columnwidth}{!}{$\displaystyle
    C_{0,0}^{\ell_{\mathrm s},\ell_{\mathrm i}}
    =\delta_{\ell_{\mathrm p},\ell_{\mathrm s}+\ell_{\mathrm i}}
    \sqrt{\frac{2}{\pi w_{\mathrm p}^2}} \frac{2^{\sigma_\ell+1}\sigma_\ell!}
    {\sqrt{|\ell_{\mathrm p}|!\,|\ell_{\mathrm s}|!\,|\ell_{\mathrm i}|!}}
    \frac{\gamma_{\mathrm s}^{|\ell_{\mathrm s}|+1}\gamma_{\mathrm i}^{|\ell_{\mathrm i}|+1}}
    {(1+\gamma_{\mathrm s}^2+\gamma_{\mathrm i}^2)^{\sigma_\ell+1}}
  $}
\end{equation*}
Yao derives this coefficient under the thin-crystal approximation from the transverse
overlap of LG modes.
Our calculation uses the full phase-matching function.
We can easily recover known results in the thin-crystal regime.
In the \(\phi\to0\) limit, Eq.~\eqref{eq:p0_unequal_waists_phase_average} gives
\begin{equation}
  C_{0,0,0}^{\ell_{\mathrm{p}},\ell_{\mathrm{s}},\ell_{\mathrm{i}}} \;
  \overset{\phi\to0}{\sim} (-1)^c\sqrt{\frac{8L}{k_{\mathrm{p}}}}\,
  C_{0,0}^{\ell_{\mathrm s},\ell_{\mathrm i}}\big|_{\mathrm{Yao}}.
  \label{eq:yao_thin_crystal_recovery}
\end{equation}
The first factor in Eq.~\eqref{eq:yao_thin_crystal_recovery} arises from the different
LG-mode phase conventions, while the second converts Yao's transverse LG-mode overlap
coefficient into the amplitude of our normalized biphoton state.
After accounting for these differences, the two results are consistent.

A Gaussian pump provides a second direct check against an established thin-crystal result.
With \(\ell_{\mathrm{p}}=0\), \(\ell_{\mathrm{s}}=\ell\), and \(\ell_{\mathrm{i}}=-\ell\),
Eq.~\eqref{eq:p0_unequal_waists_phase_average} gives
\begin{equation*}
  \frac{C_{0,0,0}^{0,\ell,-\ell}}{C_{0,0,0}^{0,0,0}} \xrightarrow[]{\phi\to0} \left(
  -\frac{2\gamma_{\mathrm{s}}\gamma_{\mathrm{i}}}
  {1+\gamma_{\mathrm{s}}^2+\gamma_{\mathrm{i}}^2} \right)^{|\ell|}.
\end{equation*}
The corresponding relative probabilities are
\(P_{\ell,-\ell}/P_{0,0} =[2\gamma_{\mathrm{s}}\gamma_{\mathrm{i}}/ (1+\gamma_{\mathrm{s}}^2+\gamma_{\mathrm{i}}^2)]^{2|\ell|}\).
This reproduces Eq.~(25) of Miatto \textit{et al.}~\cite{PhysRevA.83.033816} after
normalization to the zero-OAM coincidence probability.

We now impose equal collection waists, $w_{\mathrm{s}}=w_{\mathrm{i}}\equiv w$ and
$\gamma_{\mathrm{s}}=\gamma_{\mathrm{i}}\equiv\gamma =\frac{w_{\mathrm{p}}}{w}.$ Then
\(\rho_{\mathrm{s}}=\rho_{\mathrm{i}}=1\), \(\rho_{\mathrm{p}}=1/(2\gamma^2)\), and the
focusing relation above gives
\begin{equation}
  \phi=\phi_1 =\arctan\left(\frac{L}{2z_R^{(\mathrm{s})}}\right), \qquad \phi_2=0.
  \label{eq:symmetric_focusing_phase}
\end{equation}
Equivalently, \(\phi=\arctan(2\gamma^2\xi_{\mathrm p})\).
In this symmetric configuration, the four geometric parameters reduce to
\((\gamma,\xi_{\mathrm p})\).
We use \(\phi\) instead of \(\xi_{\mathrm p}\) to describe longitudinal focusing, giving the
parameter pair \((\gamma,\phi)\).
At fixed \(\phi\), increasing \(\gamma\) can be realized by increasing the pump waist
\(w_{\mathrm p}\) while keeping the collection waist \(w\), crystal length \(L\), and wave
numbers fixed.
This narrows the pump angular spectrum around
\(\mathbf q_{\mathrm p}=\mathbf q_{\mathrm s}+\mathbf q_{\mathrm i} \simeq\mathbf0\).
At fixed \(\gamma\), increasing the crystal length \(L\) while keeping the waists and wave
numbers fixed increases \(\phi\) and the accumulated longitudinal phase.

Equation~\eqref{eq:p0_unequal_waists_phase_average} then reduces to
\begin{equation}
  \resizebox{0.99\columnwidth}{!}{$
    \begin{aligned}
      &C^{\ell_{\mathrm{p}},\ell_{\mathrm{s}},\ell_{\mathrm{i}}}_{0,0,0}
      =\delta_{\ell_{\mathrm{p}},\ell_{\mathrm{s}}+\ell_{\mathrm{i}}} 2^{|\ell_{\mathrm{p}}|+3/2}
      \frac{2\phi}{\sqrt{\pi\tan\phi}}\, \left( \frac{\gamma}{2\gamma^2+1}
      \right)^{|\ell_{\mathrm{p}}|+1} \\
      &\times \frac{\sqrt{|\ell_{\mathrm{p}}|!\,|\ell_{\mathrm{s}}|!\,|\ell_{\mathrm{i}}|!}}
      {(|\ell_{\mathrm{s}}|-c)!\,(|\ell_{\mathrm{i}}|-c)!\,c!} \frac{1}{2\phi} \int_{-\phi}^{\phi}
      d\theta \left[ \frac{1}{2\gamma^2+1} - \cos\phi\,e^{i\theta} \right]^c
    \end{aligned}
  $}
  \label{eq:p0_dimensionless_gamma_phi_compact}
\end{equation}
For \(\ell_{\mathrm{p}}\geq0\), \(c=0\) corresponds to
\(0\leq\ell_{\mathrm{s}}\leq\ell_{\mathrm{p}}\).
For \(c=0\), the longitudinal integral is unity.
For a fixed pump OAM index, the relative amplitudes are proportional to
\(\sqrt{\ell_{\mathrm p}!/(\ell_{\mathrm s}!\,\ell_{\mathrm i}!)}\), recovering the relative probability reported in Ref.~\cite{9qrm-chgg}.

\subsection{Shaping the OAM spectrum via dimensionless source parameters}
\begin{figure*}[t]
  \centering
  \includegraphics[width=0.98\textwidth]{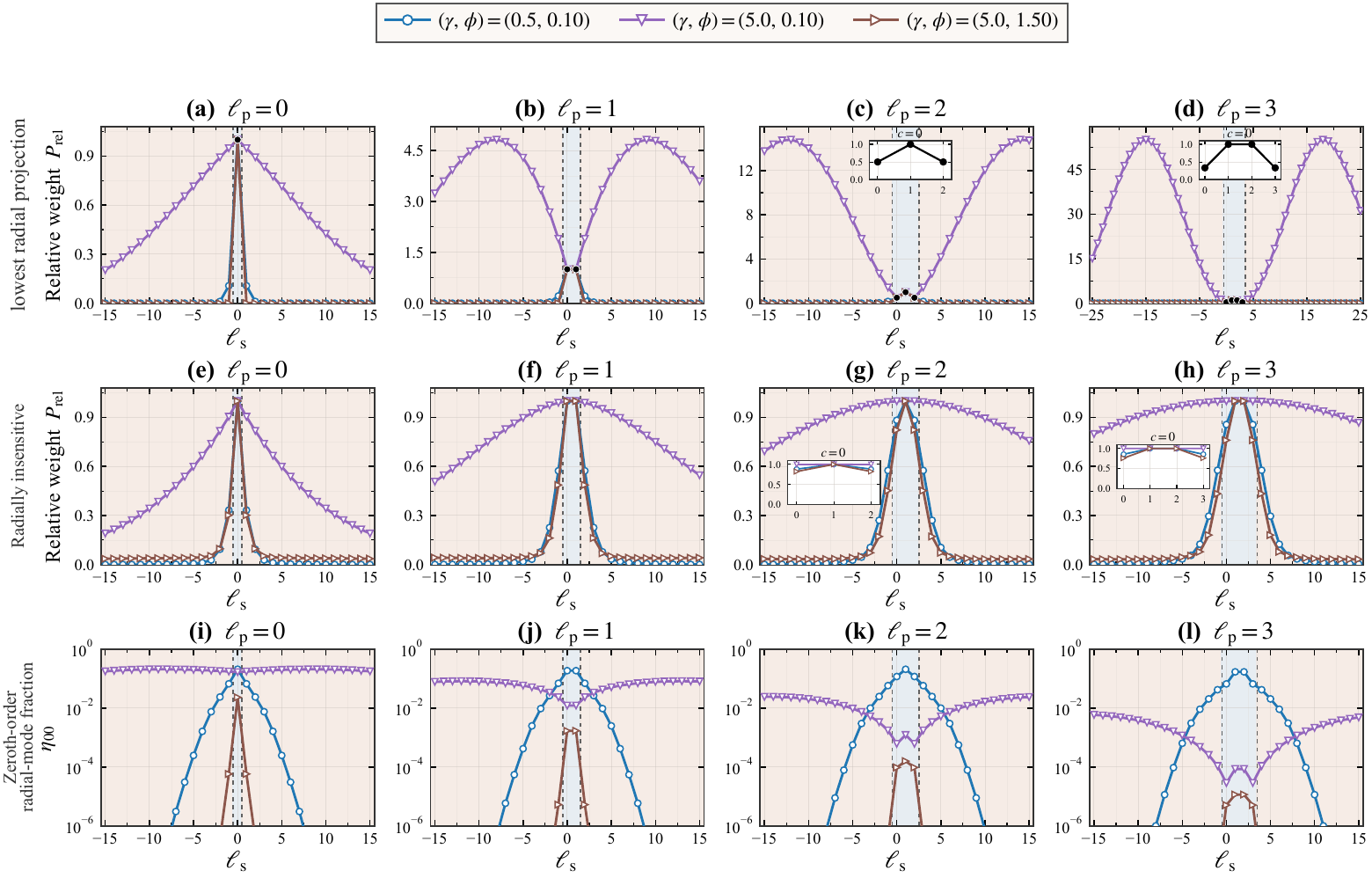}
  \caption{OAM spectra for a pump with mode indices \(p_{\mathrm{p}}=0\) and
    \(\ell_{\mathrm{p}}=0,1,2,3\) compared.
    The upper row shows the relative weight for the \(p_{\mathrm{s}}=p_{\mathrm{i}}=0\)
    projection, and the middle row shows the radially insensitive OAM spectrum
    \(P_{\mathrm{tr}}\).
    Each spectrum is divided by its maximum probability within the central interval
    $0\leq\ell_{\mathrm s}\leq\ell_{\mathrm p}$.
    The plotted $P_{\mathrm{rel}}$ is therefore a relative weight and may exceed unity outside
    this interval.
    The bottom row compares the probability ratio of both cases for each signal--idler OAM mode
    pair.
    Blue, purple, and brown lines correspond to three representative parameter pairs.
    Increasing \(\gamma\) broadens the OAM spectrum, whereas increasing \(\phi\) narrows the
    spectrum.
    The bottom row clarifies that the \(p_{\mathrm{s}}=p_{\mathrm{i}}=0\) projection accounts
    for only a small part of the available OAM distribution.
    The blue-shaded region denotes the OAM subspace
    \(0\leq\ell_{\mathrm{s}}\leq\ell_{\mathrm{p}}\) equivalently \(c=0\) and the black points in the
    insets in panels (c) and (d) show the fixed ratios that apply for all parameter pairs.
    The insets in panels (g) and (h) show the corresponding ratios, but for the radially
    insensitive spectrum the fixed ratios are lifted.}
  \label{fig:oam_sensitive_insensitive}
\end{figure*}

An OAM index $\ell$ alone does not uniquely specify a transverse mode in the LG basis, since
modes with different radial indices $p$ can carry the same OAM.
Moreover, the radial and OAM indices are generally correlated in SPDC biphoton states.
Consequently, the observed OAM spectrum depends on how the radial degree of freedom is
treated.

We compare two cases: projection onto the lowest radial order and detection without radial
postselection.
In the first case, both photons are projected onto \(p_{\mathrm s}=p_{\mathrm i}=0\), giving
the probability
$ P_{00}^{\ell_{\mathrm s},\ell_{\mathrm i}}(0,\ell_{\mathrm p}) = \left|C_{0,0,0}^{\ell_{\mathrm p},\ell_{\mathrm s},\ell_{\mathrm i}}\right|^2. $
Experimentally, this projection can be implemented using spatial light modulators and
single-mode fibers~\cite{Yao_2011,Forbes2016SLM}, without resolving higher radial modes.
The \(p_{\mathrm s}=p_{\mathrm i}=0\) subspace can contain a substantial fraction of the
generated probability, although this fraction depends strongly on the source parameters and
OAM indices, as shown below.
Its OAM spectrum follows directly from Eq.~\eqref{eq:p0_dimensionless_gamma_phi_compact}.

The radially insensitive OAM spectrum
\(P_{\mathrm{tr}}^{\ell_{\mathrm s},\ell_{\mathrm i}}(0,\ell_{\mathrm p})\) gives the total
probability of detecting a given signal--idler OAM pair irrespective of their radial
indices.
It includes contributions from all radial-mode pairs
[Eq.~\eqref{eq:radially_insensitive_oam_spectrum}] and corresponds to measurement without
radial postselection using a broadband radial detector~\cite{Karan2025BroadbandOAM}.
We evaluate the radial trace directly on the generating amplitude using the Parseval
identity and Abel limit (Appendix~\ref{app:direct-radial-trace}), avoiding individual radial
coefficient extraction and a cutoff in the radial sums.
The completeness-based formulation of Ref.~\cite{PhysRevA.97.063846} eliminates these
infinite sums analytically but still requires four-dimensional transverse integration.
Here, the transverse Gaussian integration is performed analytically before the radial trace.

Fig.~\ref{fig:oam_sensitive_insensitive} compares the two spectra for three representative
points in the $(\gamma,\phi)$ parameter space.
The upper row shows $P_{00}^{\ell_{\mathrm s},\ell_{\mathrm i}}(0,\ell_{\mathrm p})$ and the
middle row shows $P_{\mathrm{tr}}^{\ell_{\mathrm s},\ell_{\mathrm i}}(0,\ell_{\mathrm p})$.
For comparison of their shapes, each distribution is normalized to its largest probability
within the central OAM interval $0\leq\ell_{\mathrm s}\leq\ell_{\mathrm p}$.
The bottom row shows the zeroth-order radial-mode fraction
\begin{equation}
  \eta_{00}(\ell_{\mathrm s};\ell_{\mathrm p}) = \frac{
    P_{00}^{\ell_{\mathrm s},\ell_{\mathrm p}-\ell_{\mathrm s}}(0,\ell_{\mathrm p}) }{
    P_{\mathrm{tr}}^{\ell_{\mathrm s},\ell_{\mathrm p}-\ell_{\mathrm s}}(0,\ell_{\mathrm p}) },
\end{equation}
evaluated from the unscaled probabilities.
Thus, $\eta_{00}$ gives the fraction of the probability of each fixed pair of signal and
idler OAM indices that is associated with the $p_{\mathrm s}=p_{\mathrm i}=0$ projection.

The most pronounced difference between
$P_{00}^{\ell_{\mathrm s},\ell_{\mathrm i}}(0,\ell_{\mathrm p})$ and
$P_{\mathrm{tr}}^{\ell_{\mathrm s},\ell_{\mathrm i}}(0,\ell_{\mathrm p})$ occurs for
$(\gamma,\phi)=(5.0,0.10)$.
The $p_{\mathrm s}=p_{\mathrm i}=0$ OAM spectrum is centered at $\ell_{\mathrm s}=0$ for
$\ell_{\mathrm p}=0$, but shows two off-center maxima for $\ell_{\mathrm p}>0$, which move
farther from the central OAM interval as $\ell_{\mathrm p}$ increases.
In Eq.~\eqref{eq:p0_dimensionless_gamma_phi_compact}, the OAM-dependent prefactors
increasingly favor amplitudes away from the center as $\ell_{\mathrm p}$ increases, while
their suppression remains rather weak for large $\gamma$ and small $\phi$.
As a result, off-center maxima emerge.
Eq.~\eqref{eq:p0_dimensionless_gamma_phi_compact} also explains the fixed relative weights
in the insets in Figs.~\ref{fig:oam_sensitive_insensitive}(c) and
\ref{fig:oam_sensitive_insensitive}(d) for all values of $(\gamma,\phi)$ in the OAM subspace
\(0\leq\ell_{\mathrm{s}}\leq\ell_{\mathrm{p}}\), which corresponds to OAM indices where
\(c=0\).

Tracing over the radial indices changes these relative OAM weights considerably.
For the cases shown, the off-center maxima disappear and the largest probabilities occur
near $\ell_{\mathrm s}\simeq\ell_{\mathrm p}/2$.
In particular, for $(\gamma,\phi)=(5.0,0.10)$ and $\ell_{\mathrm p}>0$, the radially
insensitive OAM spectrum forms a broad, nearly uniform plateau extending several modes
beyond the central OAM interval.
This demonstrates that the higher-radial-order contributions are not distributed uniformly
over OAM and can therefore qualitatively reshape the OAM spectrum.
In the insets of (g) and (h) we clearly recognize that the fixed relative ratios no longer apply.

The bottom row quantifies the contribution of the \(p_{\mathrm s}=p_{\mathrm i}=0\)
projection to the radially insensitive OAM spectrum.
For $(\gamma,\phi)=(5.0,0.10)$, $\eta_{00}$ has a minimum near
$\ell_{\mathrm s}\simeq\ell_{\mathrm p}/2$ and increases toward both sides.
For $(0.5,0.10)$, both spectra decrease toward OAM modes with larger $|\ell_{\mathrm s}|$,
but the $p_{\mathrm s}=p_{\mathrm i}=0$ OAM spectrum decreases more rapidly.
For $(5.0,1.50)$, the contribution from $p_{\mathrm s}=p_{\mathrm i}=0$ decreases strongly
toward larger $|\ell_{\mathrm s}|$, while the radially insensitive OAM spectrum remains
non-negligible.
Thus, for these OAM components, most of the probability originates from higher radial
indices.
Increasing $\gamma$ at $\phi=0.10$ changes the zeroth-order radial-mode fraction from a
central peak to maxima farther from the central OAM interval, whereas at $\gamma=5$,
increasing $\phi$ restores a central maximum but shrinks this fraction overall.

The broadening with \(\gamma\) does not contradict the narrowing of the pump angular
spectrum.
A larger pump waist strengthens the transverse wave-vector anticorrelation
\(\mathbf q_{\mathrm s}\simeq-\mathbf q_{\mathrm i}\), but this constraint acts on the sum
of the transverse wave vectors rather than on either OAM index separately.
It can therefore support a wider range of signal and idler OAM indices that satisfy
\(\ell_{\mathrm p}=\ell_{\mathrm s}+\ell_{\mathrm i}\).
By contrast, a larger \(\phi\) produces stronger longitudinal phase variation across the
crystal.
Contributions with larger transverse phase mismatch then undergo stronger cancellation in
the longitudinal integral, which suppresses the relative weight of OAM modes with large
absolute indices under the conditions compared in Fig.~\ref{fig:oam_sensitive_insensitive}.

\subsection{Dimensional scaling of target-state fidelities}
Tuning the source parameters can produce nearly uniform OAM spectra, but OAM probabilities
do not specify the coherences required for an MES.
We therefore study the fidelity with a $d$-dimensional MES target as a function of $d$,
comparing the $p_{\mathrm s}=p_{\mathrm i}=0$ projection with the radially insensitive OAM
state.
The experimentally accessible OAM dimensionality is constrained by both the source geometry
and the measurement bandwidth of the mode-projection
system~\cite{PhysRevA.83.033816,Miatto2012OAMBandwidth}.

For both detection schemes, $S_{\ell_{\mathrm p},d}$ contains the $d$ signal OAM indices in
the target, with idler indices fixed by OAM conservation as
$\ell_{\mathrm i}=\ell_{\mathrm p}-\ell$.
For the nonnegative pump OAM indices considered here, we choose
\(S_{\ell_{\mathrm p},d}=\{-k,\ldots,\ell_{\mathrm p}+k\}\), with
\(d=|\ell_{\mathrm p}|+1+2k\).
In particular, $S_{0,d}=\{-k,\ldots,k\}$.
The global fidelity measures the overlap of the full generated state with the target,
whereas the subspace fidelity measures the overlap after projection onto the target subspace
and renormalization.
The global fidelity therefore equals the subspace fidelity multiplied by the probability of
occupying the corresponding target subspace.

For the $p_{\mathrm s}=p_{\mathrm i}=0$ projection, the equal-amplitude MES target of $d$
full LG modes is
\begin{equation}
  \ket{\Phi^{(00)}_{\ell_{\mathrm p},d}} = \frac{1}{\sqrt{d}}
  \sum_{\ell\in S_{\ell_{\mathrm p},d}} \ket{0,\ell}_{\mathrm s}
  \ket{0,\ell_{\mathrm p}-\ell}_{\mathrm i}.
  \label{eq:lg_bell_state_00}
\end{equation}
For the full generated state $\ket{\psi}$ of Eq.~\eqref{eq:biphoton_lg_expansion}, the
global fidelity is
\begin{equation}
  \makebox[0.8\columnwidth][c]{$\displaystyle
    F_{\mathrm{global,sens}}^{(\ell_{\mathrm p},d)}
    = |\braket{\psi|\Phi^{(00)}_{\ell_{\mathrm p},d}}|^2  = \frac{1}{d} \left|
    \sum_{\ell\in S_{\ell_{\mathrm p},d}}
    C^{\ell_{\mathrm p},\ell,\ell_{\mathrm p}-\ell}_{0,0,0} \right|^2 .
  $}
  \label{eq:global_bell_fidelity_general}
\end{equation}
Conditioning on this LG subspace gives the subspace fidelity
\begin{equation}
  F_{\mathrm{sub,sens}}^{(\ell_{\mathrm p},d)} = \frac{ \left|
    \displaystyle\sum_{\ell\in S_{\ell_{\mathrm p},d}}
    C^{\ell_{\mathrm p},\ell,\ell_{\mathrm p}-\ell}_{0,0,0} \right|^2 }{
    d\displaystyle\sum_{\ell\in S_{\ell_{\mathrm p},d}} \left|
    C^{\ell_{\mathrm p},\ell,\ell_{\mathrm p}-\ell}_{0,0,0} \right|^2 }.
  \label{eq:subspace_fidelity_00}
\end{equation}

For the radially insensitive case, tracing over the signal and idler radial indices gives
the reduced OAM state
\(\rho_{\mathrm{OAM}} =\operatorname{Tr}_{p_{\mathrm s},p_{\mathrm i}} (\ket{\psi}\bra{\psi})\).
The corresponding MES target in the reduced OAM Hilbert space is
\begin{equation}
  \ket{\Phi^{(\mathrm{OAM})}_{\ell_{\mathrm p},d}} = \frac{1}{\sqrt{d}}
  \sum_{\ell\in S_{\ell_{\mathrm p},d}} \ket{\ell}_{\mathrm s}
  \ket{\ell_{\mathrm p}-\ell}_{\mathrm i}
\end{equation}
The global fidelity is
\begin{equation}
  \begin{aligned}
    F_{\mathrm{global,ins}}^{(\ell_{\mathrm p},d)}
    & = \bra{\Phi^{(\mathrm{OAM})}_{\ell_{\mathrm p},d}} \rho_{\mathrm{OAM}}
    \ket{\Phi^{(\mathrm{OAM})}_{\ell_{\mathrm p},d}} \\
    &= \frac{1}{d} \sum_{p_{\mathrm s},p_{\mathrm i}=0}^{\infty} \left|
    \sum_{\ell\in S_{\ell_{\mathrm p},d}}
    C^{\ell_{\mathrm p},\ell,\ell_{\mathrm p}-\ell}_{0,p_{\mathrm s},p_{\mathrm i}} \right|^2 .
  \end{aligned}
  \label{eq:fidelity_trace_global}
\end{equation}
The subspace fidelity is
\begin{equation}
  F_{\mathrm{sub,ins}}^{(\ell_{\mathrm p},d)} = \frac{
    \displaystyle\sum_{p_{\mathrm s},p_{\mathrm i}=0}^{\infty} \left|
    \displaystyle\sum_{\ell\in S_{\ell_{\mathrm p},d}}
    C^{\ell_{\mathrm p},\ell,\ell_{\mathrm p}-\ell}_{0,p_{\mathrm s},p_{\mathrm i}} \right|^2 }{
    d\displaystyle\sum_{p_{\mathrm s},p_{\mathrm i}=0}^{\infty}
    \sum_{\ell\in S_{\ell_{\mathrm p},d}} \left|
    C^{\ell_{\mathrm p},\ell,\ell_{\mathrm p}-\ell}_{0,p_{\mathrm s},p_{\mathrm i}} \right|^2 }.
  \label{eq:subspace_fidelity_trace}
\end{equation}
In Eqs.~\eqref{eq:fidelity_trace_global} and \eqref{eq:subspace_fidelity_trace}, OAM
amplitudes are summed coherently within each radial-mode pair
$(p_{\mathrm s},p_{\mathrm i})$, and the resulting probabilities are combined incoherently
over radial-mode pairs.

The method of Ref.~\cite{Karan2025BroadbandOAM} measures OAM spectra without radial-mode
postselection but does not access the coherences needed to determine MES fidelities.
The fidelities defined here refer to the reduced OAM state obtained by tracing over the LG
radial indices.
Therefore, experimentally determining these fidelities requires additional measurements in
OAM superposition bases, with the same relative amplitudes and phases in the measurement
projections for all contributing radial-mode pairs.

\begin{figure*}[t]
  \centering
  \includegraphics[width=\textwidth]{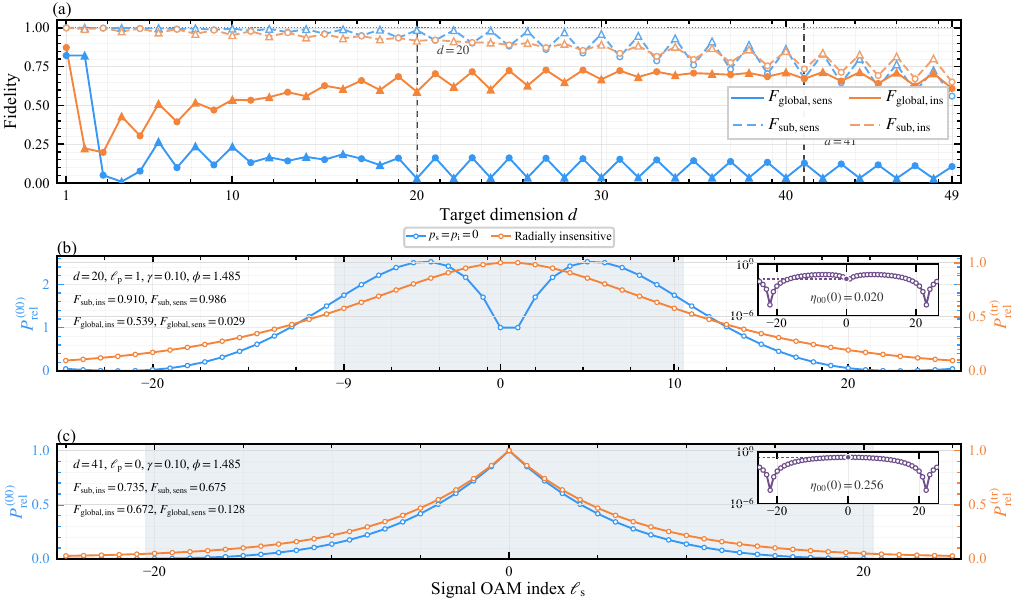}
  \captionsetup{justification=justifiedhyphen,singlelinecheck=false}
  \caption{Dimensional scaling of the fidelity with maximally entangled target states. (a)
    Global (solid) and best-found subspace (dashed) fidelities for $d=1,\ldots,49$.
    Circles denote odd $d$ with $\ell_{\mathrm p}=0$, and triangles even $d$ with
    $\ell_{\mathrm p}=1$.
    Blue denotes the $p_{\mathrm s}=p_{\mathrm i}=0$ projection and orange the radially
    insensitive OAM state.
    For each $d$ and detection scheme, the global fidelity uses the same parameters as the best
    subspace fidelity found within the ranges specified in the text.
    Vertical dashed lines mark $d=20$ and $41$. (b,c) Corresponding OAM spectra for these
    dimensions, with shaded target subspaces.
    As in Fig.~\ref{fig:oam_sensitive_insensitive}, the insets show the zeroth-order radial-mode
    fraction $\eta_{00}$ for each signal--idler OAM pair.}
  \label{fig:trace_fidelity}
\end{figure*}

We choose the search ranges $0.1\leq\gamma\leq10$ and $0.1\leq\phi\leq1.5$ with experimental
feasibility in mind.
For each target dimension and detection scheme, we select the best subspace fidelity found
within these ranges and evaluate the global fidelity at the same parameter pair
$(\gamma,\phi)$.
We use $\ell_{\mathrm p}=0$ for odd target dimensions and $\ell_{\mathrm p}=1$ for even
dimensions.
Fig.~\ref{fig:trace_fidelity}(a) shows an overall decrease in the best-found subspace
fidelity with increasing $d$ for both schemes.
Maintaining the required coherent superposition over more OAM indices becomes increasingly
difficult within these parameter ranges.

For the $p_{\mathrm s}=p_{\mathrm i}=0$ projection, high subspace fidelity can coexist with
low global fidelity because only a small fraction of the generated probability occupies the
selected LG subspace.
This is particularly pronounced for even-dimensional targets with $\ell_{\mathrm p}=1$,
whose global fidelity is close to zero.
For $\ell_{\mathrm p}=0$, the global fidelity instead peaks at $d=23$.
Initially, expanding the target around the central state captures more probability despite
the decreasing subspace fidelity.
Beyond $d=23$, the additional probability no longer compensates for this decrease.

For the radially insensitive OAM state, expanding the target subspace captures more of the
spectrum and reduces the gap between global and subspace fidelity.
This effect is strongest for $\ell_{\mathrm p}=0$, where the subspace expands symmetrically
around a single central state.
For $\ell_{\mathrm p}=1$, the spectrum is centered on the two OAM mode pairs
$(\ell_{\mathrm s},\ell_{\mathrm i})=(0,1)$ and $(1,0)$ and has broader outer contributions,
so the two fidelities approach each other more slowly.
Greater occupation does not guarantee high subspace fidelity because different radial-mode
pairs need not contribute the same coherent OAM superposition.
The best-found subspace fidelity still decreases with $d$.
Pump OAM indices $\ell_{\mathrm p}=2$ and $3$ provided no improvement in dimensionality
scaling over $0$ and $1$ within the same parameter ranges.

Fig.~\ref{fig:trace_fidelity}(b,c) shows the spectra for $d=20$ and $41$ at the best-found
parameters $(\gamma,\phi)=(0.10,1.485)$, with $\ell_{\mathrm p}=1$ and $0$, respectively.
For $d=20$, the $p_{\mathrm s}=p_{\mathrm i}=0$ projection gives a high subspace fidelity,
but its target subspace contains only $2.9\%$ of the generated probability, resulting in
$F_{\mathrm{global,sens}}=0.029$.
The radially insensitive state has a much larger target-subspace occupation of $59.2\%$, as
the long spectral tails contribute substantial probability across the target range despite
the nonuniform OAM distribution.
This keeps $F_{\mathrm{global,ins}}$ closer to $F_{\mathrm{sub,ins}}$.
For $d=41$, the corresponding occupations are $19.0\%$ and $91.4\%$, while both subspace
fidelities are lower than for $d=20$.
The insets show the fraction $\eta_{00}$ associated with $p_{\mathrm s}=p_{\mathrm i}=0$ for
each OAM pair.
Its strong variation shows that the radial composition depends on the OAM indices.
Although the spectra are visibly nonuniform, particularly at $d=20$, the amplitude
magnitudes for the $p_{\mathrm s}=p_{\mathrm i}=0$ projection scale as
$\sqrt{P_{\mathrm{rel}}}$ and are therefore more uniform than the plotted probabilities.

\begin{figure*}[t]
  \centering
  \includegraphics[width=\textwidth]{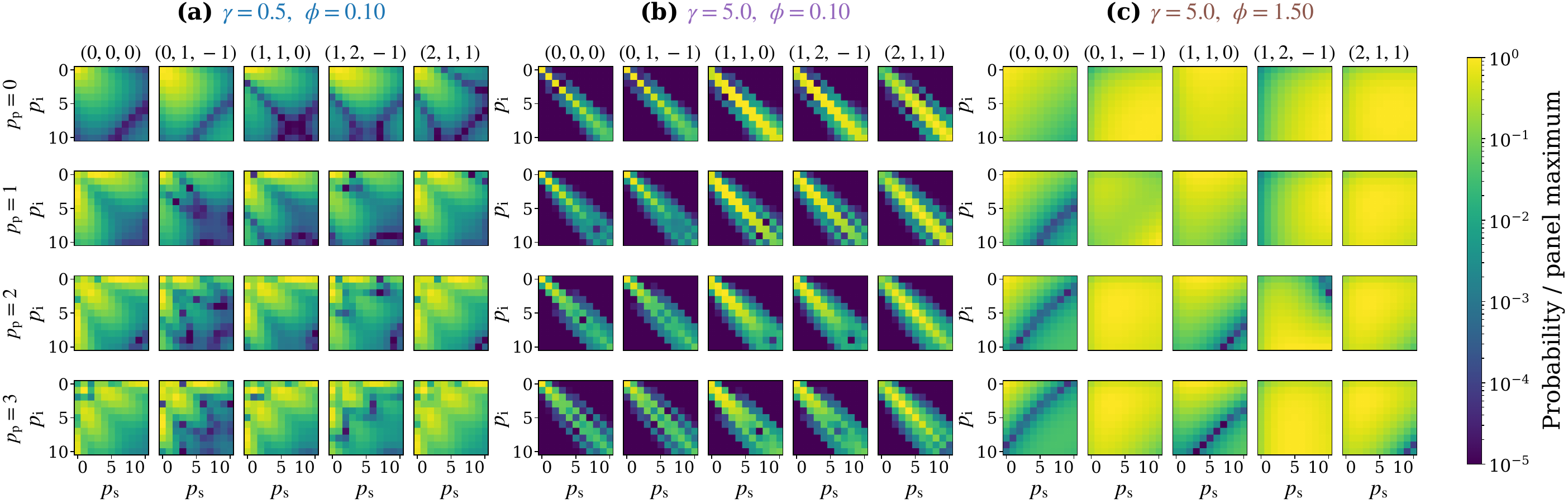}
  \caption{Comparison of radial-mode
    distributions for selected OAM-index triples.
    Panels (a)--(c) correspond to the same $(\gamma,\phi)$ parameter pairs as in
    Fig.~\ref{fig:oam_sensitive_insensitive}.
    Rows give the pump radial index $p_{\mathrm{p}}=0,1,2,3$, and the columns show selected
    OAM-index triples $(\ell_{\mathrm{p}},\ell_{\mathrm{s}} ,\ell_{\mathrm{i}})$.
    Each heat map shows
    $|C_{p_{\mathrm{p}},p_{\mathrm{s}},p_{\mathrm{i}}}^{\ell_{\mathrm{p}},\ell_{\mathrm{s}},\ell_{\mathrm{i}}}|^2$
    for $p_{\mathrm{s}},p_{\mathrm{i}}=0,\ldots,10$, divided by the maximum probability within
    that $11\times11$ panel to compare the radial-distribution shapes.
    At small \(\phi\), increasing \(\gamma\) changes broad and diffuse radial distributions into
    a pronounced ridge near \(p_{\mathrm{s}}\simeq p_{\mathrm{i}}\) for all OAM indices.
    Increasing \(\phi\) at fixed large \(\gamma\) replaces this ridge with a broader radial-mode
    distribution.
    Interestingly, for $p_{\mathrm{p}}>0$ we see the formation of anti-diagonal low-probability
    troughs for certain OAM indices, which indicate strongly suppressed detection probabilities.
  }
  \label{fig:radial-structure-c0}
\end{figure*}

High-dimensional MES generation requires both high subspace fidelity and substantial
occupation of that subspace, both of which depend on the treatment of radial modes.
Varying $\gamma$ and $\phi$ alone can achieve high subspace fidelities at small $d$, but at
the expense of low global fidelity.
Further amplitude control is offered by radial pump-mode superpositions for
postselection-free OAM-spectrum shaping~\cite{Karan2023PostselectionFree}, spatial pump
engineering combined with path identity~\cite{9qrm-chgg}, and modal-space design through
postselection~\cite{Dada2011HighDimOAM}.

\subsection{Characterization of the radial-mode distribution}

The signal and idler OAM indices $(\ell_{\mathrm s},\ell_{\mathrm i})$ do not uniquely
specify an LG biphoton state.
Probability can be distributed over many radial-mode pairs $(p_{\mathrm s},p_{\mathrm i})$,
and the measured probability depends on the chosen radial projection.
Predicting radial correlations is harder because no analogous selection rule fixes the
radial indices.
Using interferometric imaging, Zia \textit{et al.}~\cite{Zia2023Interferometric}
reconstructed the amplitude and phase of spatial biphoton states.
For $(\ell_{\mathrm s},\ell_{\mathrm i})=(\ell_{\mathrm p},0)$ and matched pump and
decomposition waists, the radial correlations peaked when one photon had radial index
$p_{\mathrm p}$ and the other zero, directly linking the pump radial order to the
signal--idler correlations.
Full LG-mode measurements with structured pumps at $p_{\mathrm p}=0,1,2,3$ and
$\ell_{\mathrm p}=0,1,2$ showed strong diagonal radial correlations at low pump radial
orders and increasing off-diagonal contributions for higher pump orders and OAM
subspaces~\cite{DErrico2021FullMode}.
Radial-mode sorters further enable $p$-resolved measurements and the use of radial
modes~\cite{Zhou2017RadialSorter,Gu2018RadialSorter}.
Keeping the full phase-matching function, we extend this characterization beyond the
thin-crystal regime and examine how the pump radial order, OAM indices, and parameters
$\gamma$ and $\phi$ affect the coincidence amplitudes and joint radial-mode distributions.

Fig.~\ref{fig:radial-structure-c0} uses the same three $(\gamma,\phi)$ parameter pairs as
Fig.~\ref{fig:oam_sensitive_insensitive}.
Rows give $p_{\mathrm p}=0,1,2,3$, and columns give the OAM-index triples
$(\ell_{\mathrm p},\ell_{\mathrm s},\ell_{\mathrm i})=(0,0,0)$, $(0,1,-1)$, $(1,1,0)$,
$(1,2,-1)$, and $(2,1,1)$.
Each $11\times11$ heat map shows
$\left|C_{p_{\mathrm p},p_{\mathrm s},p_{\mathrm i}} ^{\ell_{\mathrm p},\ell_{\mathrm s},\ell_{\mathrm i}}\right|^2$
for $p_{\mathrm s},p_{\mathrm i}=0,\ldots,10$, divided by its maximum probability.

At fixed $\phi=0.10$, panels (a) and (b) show how increasing $\gamma$ strengthens the
diagonal radial correlation.
For the narrow pump in panel (a), probability spreads over both diagonal and off-diagonal
radial-mode pairs.
For the broad pump in panel (b), it concentrates along $p_{\mathrm s}\simeq p_{\mathrm i}$,
forming a pronounced ridge for all selected OAM configurations.

At fixed small \(\phi\), a larger \(\gamma\) narrows the pump angular spectrum, enforcing
\(\mathbf q_{\mathrm s}+\mathbf q_{\mathrm i}\simeq\mathbf 0\) more strongly.
Projection onto the matched signal and idler LG bases then favors similar radial orders,
producing the ridge near \(p_{\mathrm s}\simeq p_{\mathrm i}\).

At fixed $\gamma=5.0$, increasing $\phi$ from $0.10$ in panel (b) to $1.50$ in panel (c)
removes the diagonal ridge and broadens the radial distributions, revealing a stronger
dependence on the pump radial order.
For $p_{\mathrm p}=0$, the distributions remain comparatively smooth over the displayed
range.
At higher $p_{\mathrm p}$, mode-dependent low-probability troughs appear, most prominently
for $(0,0,0)$ and $(1,1,0)$, and move toward the matrix center as $p_{\mathrm p}$ increases.

The troughs arise from coherent cancellation in the coincidence amplitude.
Pump modes with \(p_{\mathrm p}>0\) have radial nodes and lobes with alternating phase.
For particular pairs \((p_{\mathrm s},p_{\mathrm i})\), contributions from different radial
regions and longitudinal positions cancel before the amplitude is squared.
Increasing \(\phi\) changes the longitudinal phase weighting, making these cancellations
more visible as the diagonal correlation weakens.
The trough locations depend on the pump radial order and selected OAM indices, so they
should not be interpreted as an additional conservation rule.

When the strong $p_{\mathrm s}\simeq p_{\mathrm i}$ correlation dominates, the radial
distributions are nearly independent of the selected OAM-index triples.
As this correlation weakens, different OAM-index triples lead to increasingly different
radial distributions.

Salakhutdinov \textit{et al.}~\cite{PhysRevLett.108.173604} observed strong diagonal radial
correlations with a reported detection-mode waist of $500\,\mu\mathrm m$ at the SLM,
corresponding to $66.7\,\mu\mathrm m$ at the crystal.
Using this value with their experimental parameters and $n_{\mathrm p}=1.85$ for KTP gives
$\gamma=4.88$ and $\phi=0.032$, placing the experiment in the same large-$\gamma$,
small-$\phi$ regime as panel (b).
The $p_{\mathrm s}\simeq p_{\mathrm i}$ correlation is pronounced only in this regime.
Increasing $\phi$ at fixed large $\gamma$ weakens it and produces more complex distributions
with stronger dependence on the pump radial order and OAM indices.
To obtain a strongly diagonal radial-mode distribution, experiments should keep $\phi$ small
and tune $\gamma$ within the large-waist-ratio regime toward the diagonal-correlation
maximum.

All 60 of the $11\times11$ radial spectra in Fig.~\ref{fig:radial-structure-c0} were
evaluated in approximately 14 seconds, enabling systematic exploration across experimentally
relevant pump and OAM indices and direct predictions for experiments.

\section{Conclusion}
Our generating-function approach calculates LG coincidence amplitudes for arbitrary pump,
signal, and idler LG indices without approximating the PMF.
The beam waists, wave numbers, and crystal length enter through four dimensionless
parameters \((\gamma_{\mathrm s},\gamma_{\mathrm i},\xi_{\mathrm p},\nu)\), which reduce to
\((\gamma,\phi)\) for wave-vector-degenerate SPDC with equal signal and idler waists.
The lowest-radial-order amplitudes recover established analytical expressions in the
thin-crystal limit.
Higher radial orders are evaluated through a compact expansion with finite coefficients,
yielding numerically stable results within one second in the benchmark cases.
We also obtain a convergent evaluation of the infinite sum over \(p_{\mathrm s}\) and
\(p_{\mathrm i}\) for radially insensitive OAM detection.

We use these source parameters to shape LG spectra and improve fidelities with MES targets
under different radial detection schemes.
Increasing \(\gamma\) enhances the relative probabilities of OAM modes with large absolute
indices, whereas increasing \(\phi\) suppresses them.
The treatment of radial modes determines the detailed spectral profile.
High-dimensional MES generation requires both high fidelity within the target OAM subspace
and substantial occupation of that subspace.
The \(p_{\mathrm s}=p_{\mathrm i}=0\) projection can yield high conditional MES fidelity but
low preparation efficiency, whereas radially insensitive OAM detection collects more of the
generated spectrum but combines radial contributions incoherently.
Within the explored parameter ranges, the two schemes give subspace fidelities of $0.986$
and $0.910$ at $d=20$, with target-subspace occupations of $0.029$ and $0.592$,
respectively.

The radial-mode distributions show that increasing \(\gamma\) concentrates probability along
\(p_{\mathrm s}\simeq p_{\mathrm i}\), while increasing \(\phi\) replaces this diagonal structure with nearly
uniform distributions.
These results guide the selection of source parameters for studying and engineering
high-dimensional spatial entanglement.

Tailored pump-mode superpositions~\cite{Bernecker2024MaximallyEntangled} and patterned
nonlinear crystals with a nonuniform longitudinal profile
$\widetilde{\chi}(t)$~\cite{Baghdasaryan2023DomainEngineering} generally increase the
complexity of LG coincidence-amplitude calculations.
Our approach may provide a starting point for their treatment at feasible computational
cost.
The dimensionless formulation can further simplify source optimization by identifying
independent geometric controls and physically distinct configurations that yield the same
normalized LG spectrum.
Extending the present spectrally narrowband framework to broadband SPDC requires explicitly
including the frequency dependence of the wave vectors and phase mismatch.
Such an extension could describe how pump bandwidth, material dispersion, and spectral
filtering shape the full LG spectrum through spatial--spectral coupling.

\begin{acknowledgments}
  J.P. and R.B. acknowledge support from the Research School of Advanced Photon Science of the
  Helmholtz Institute Jena.
  This work was supported by the Honours Programme for Future Researchers at Friedrich
  Schiller University Jena (project HPJ-UP51).

\end{acknowledgments}

\appendix

\section{Details of the analytical reduction}
\label{app:derivation}
This appendix gives the explicit coefficients and the radial expansion underlying the
reduction in Sec.~II\,C.
We use \(t\) and \(\widetilde\chi(t)\) as defined in
Eq.~\eqref{eq:dimensionless_longitudinal_coordinate}, with the constant nonlinear scale
absorbed into the global normalization.

\subsection{Gaussian transverse integral}
\label{app:gaussian-integral}
The coordinates \(\mathbf q_\pm=\mathbf q_{\mathrm s}\pm\mathbf q_{\mathrm i}\) used in the
transverse integration give \(\mathbf q_{\mathrm s}=(\mathbf q_++\mathbf q_-)/2\) and
\(\mathbf q_{\mathrm i}=(\mathbf q_+-\mathbf q_-)/2\), with
\(d^2\mathbf q_{\mathrm s}\,d^2\mathbf q_{\mathrm i} =\frac14\,d^2\mathbf q_+\,d^2\mathbf q_-\).
The coefficients of the phase quadratic form are
\begin{equation*}
  \begin{aligned}
    \zeta_+&=\frac18\left(\frac1{k_{\mathrm s}}+\frac1{k_{\mathrm i}}\right)
    -\frac1{2k_{\mathrm p}},\\
    \zeta_-&=\frac18\left(\frac1{k_{\mathrm s}}+\frac1{k_{\mathrm i}}\right),\qquad
    \zeta_{+-}=\frac14\left(\frac1{k_{\mathrm s}}-\frac1{k_{\mathrm i}}\right).
  \end{aligned}
\end{equation*}
For \(j\in\{\mathrm p,\mathrm s,\mathrm i\}\), define
\begin{equation*}
  \beta_j(\eta_j)=\frac{w_j^2(1-\eta_j)}{4(1+\eta_j)},\qquad \mathbf h_j(\eta_j,\mu_j)
  =\frac{iw_j\mu_j}{2(1+\eta_j)}(1,i\sigma_j).
\end{equation*}
Here \(\sigma_j\in\{+1,-1\}\) is the sign of \(\ell_j\) for nonzero OAM.
For \(\ell_j=0\), either choice gives the same extracted mode.
The stars on \(\mathbf h_{\mathrm s}\) and \(\mathbf h_{\mathrm i}\) follow the
coefficientwise conjugation convention of Eq.~\eqref{eq:generating_amplitude}, while
\(\beta_j\) has real Taylor coefficients.
Suppressing function arguments, the Gaussian coefficients are
\begin{equation*}
  \begin{gathered}
    A_+=\beta_{\mathrm p}+\frac{\beta_{\mathrm s}+\beta_{\mathrm i}}4 +it\frac L2\zeta_+,\quad
    A_-=\frac{\beta_{\mathrm s}+\beta_{\mathrm i}}4+it\frac L2\zeta_-,\\
    A_{+-}=\frac{\beta_{\mathrm s}-\beta_{\mathrm i}}2+it\frac L2\zeta_{+-},\\
    \mathbf J_+=\mathbf h_{\mathrm p} +\frac{\mathbf h_{\mathrm s}^*
      +\mathbf h_{\mathrm i}^*}2,\qquad \mathbf J_-=\frac{\mathbf h_{\mathrm s}^*
      -\mathbf h_{\mathrm i}^*}2.
  \end{gathered}
\end{equation*}
Together with \(\Delta=A_+A_--A_{+-}^2/4\), these give
Eq.~\eqref{eq:transverse_gaussian_result}.
For \(|\eta_j|<1\), \(\operatorname{Re}\beta_j=w_j^2(1-|\eta_j|^2)/(4|1+\eta_j|^2)>0\).
The real part of the Gaussian quadratic form is therefore positive definite, ensuring
absolute convergence near \(\boldsymbol\eta=\mathbf0\).

\subsection{Angular coefficient extraction and OAM conservation}
\label{app:angular-extraction}
The coefficients in Eq.~\eqref{eq:ABC_angular_expansion} are
\begin{equation}
  \begin{aligned}
    A&=\frac{w_{\mathrm p}w_{\mathrm s}(A_--A_{+-}/2)}
    {8\Delta(1+\eta_{\mathrm p})(1+\eta_{\mathrm s})}
    \delta_{\sigma_{\mathrm p},\sigma_{\mathrm s}},\\
    B&=\frac{w_{\mathrm p}w_{\mathrm i}(A_-+A_{+-}/2)}
    {8\Delta(1+\eta_{\mathrm p})(1+\eta_{\mathrm i})}
    \delta_{\sigma_{\mathrm p},\sigma_{\mathrm i}},\\
    C&=-\frac{w_{\mathrm s}w_{\mathrm i}(A_--A_+)}
    {16\Delta(1+\eta_{\mathrm s})(1+\eta_{\mathrm i})}
    \delta_{\sigma_{\mathrm s},-\sigma_{\mathrm i}}.
  \end{aligned}
  \label{app:ABC}
\end{equation}
Their sign constraints, together with the nonnegative integer powers in
Eq.~\eqref{eq:angular_integers_ab}, give the OAM selection rule derived in the main text.

\subsection{Longitudinal kernel}
\label{app:longitudinal-kernel}
At \(\boldsymbol\eta=\mathbf0\), the determinant factorizes as
\begin{equation}
  \begin{aligned}
    \Delta(t;\mathbf0) &=\frac{w_{\mathrm p}^4}{64}(D+iEt-Ft^2)\\
    &=\frac{w_{\mathrm p}^4D}{64} (1+it\tan\phi_1)(1+it\tan\phi_2),
  \end{aligned}
  \label{app:eq:determinant-factorization}
\end{equation}
with
\begin{equation*}
  \begin{aligned}
    D&=\frac1{\gamma_{\mathrm s}^2}+\frac1{\gamma_{\mathrm i}^2}
    +\frac1{\gamma_{\mathrm s}^2\gamma_{\mathrm i}^2},\\
    E&=\xi_{\mathrm p}\biggl[
    \left(1+\frac1{\gamma_{\mathrm i}^2}\right)\frac{k_{\mathrm p}}{k_{\mathrm s}}
    +\left(1+\frac1{\gamma_{\mathrm s}^2}\right)\frac{k_{\mathrm p}}{k_{\mathrm i}}
    -\frac1{\gamma_{\mathrm s}^2}-\frac1{\gamma_{\mathrm i}^2} \biggr],\\
    F&=\xi_{\mathrm p}^2\left[ \frac{k_{\mathrm p}^2}{k_{\mathrm s}k_{\mathrm i}}
    -\frac{k_{\mathrm p}}{k_{\mathrm s}}-\frac{k_{\mathrm p}}{k_{\mathrm i}} \right].
  \end{aligned}
\end{equation*}
The roots give the effective focusing parameters in
Eq.~\eqref{eq:effective_focusing_phases}.
Combining this factorization with the numerator factors from \(A^aB^bC^c\) gives the
longitudinal kernel in Eq.~\eqref{eq:oam_gouy_kernel}, after all integration-independent
factors are removed.

\subsection{Radial coefficient extraction}
\label{app:radial-extraction}
The exponent in Eq.~\eqref{eq:radial_coefficient_extraction} is given below, together with
the radial correction \(\mathcal Q_\eta\) to the coupled denominator.
\begin{widetext}
  \begin{equation}
    \begin{aligned}
      \Psi(t;\boldsymbol\eta) ={}&-(a+b+c+1)\ln\!\left[
      (1+it\tan\phi_1)(1+it\tan\phi_2)+\frac{\mathcal Q_\eta}{D}\right]\\
      &-(a+b+1)\ln(1+\eta_{\mathrm p}) -(a+c+1)\ln(1+\eta_{\mathrm s})
      -(b+c+1)\ln(1+\eta_{\mathrm i})\\
      &+a\ln\!\left(1+it\tan\phi_{\mathrm i} -\frac{2\eta_{\mathrm i}}{1+\eta_{\mathrm i}}\right)
      +b\ln\!\left(1+it\tan\phi_{\mathrm s} -\frac{2\eta_{\mathrm s}}{1+\eta_{\mathrm s}}\right)
      +c\ln\!\left(-1+it\tan\phi_{\mathrm p} +\frac{2\eta_{\mathrm p}}{1+\eta_{\mathrm p}}\right).
    \end{aligned}
    \label{Psi}
  \end{equation}
  \begin{equation*}
    \begin{aligned}
      \mathcal Q_\eta ={}&\frac1{\gamma_{\mathrm s}^2}\left[
      \frac{1-\eta_{\mathrm s}}{1+\eta_{\mathrm s}}
      \frac{1-\eta_{\mathrm p}}{1+\eta_{\mathrm p}}-1\right] +\frac1{\gamma_{\mathrm i}^2}\left[
      \frac{1-\eta_{\mathrm i}}{1+\eta_{\mathrm i}}
      \frac{1-\eta_{\mathrm p}}{1+\eta_{\mathrm p}}-1\right]
      +\frac1{\gamma_{\mathrm s}^2\gamma_{\mathrm i}^2}\left[
      \frac{1-\eta_{\mathrm s}}{1+\eta_{\mathrm s}}
      \frac{1-\eta_{\mathrm i}}{1+\eta_{\mathrm i}}-1\right]\\
      &+it\frac L{w_{\mathrm p}^2}\biggl\{
      \left(\frac1{k_{\mathrm s}}+\frac1{k_{\mathrm i}}\right)
      \left[\frac{1-\eta_{\mathrm p}}{1+\eta_{\mathrm p}}-1\right]
      +\frac1{\gamma_{\mathrm s}^2}\left(\frac1{k_{\mathrm i}}-\frac1{k_{\mathrm p}}\right)
      \left[\frac{1-\eta_{\mathrm s}}{1+\eta_{\mathrm s}}-1\right]
      +\frac1{\gamma_{\mathrm i}^2}\left(\frac1{k_{\mathrm s}}-\frac1{k_{\mathrm p}}\right)
      \left[\frac{1-\eta_{\mathrm i}}{1+\eta_{\mathrm i}}-1\right] \biggr\}.
    \end{aligned}
  \end{equation*}
\end{widetext}
Since \(\mathcal Q_\eta=0\) at \(\boldsymbol\eta=\mathbf0\), \(\Psi(t;\mathbf0)\) contains
the lowest-radial-order contribution removed in
Eq.~\eqref{eq:radial_coefficient_extraction}.

\subsection{Combinatorial expansion of radial coefficients}
\label{app:rho-expansion}
Taylor expansion about \(\boldsymbol\eta=\mathbf0\) gives
\begin{equation}
  \Psi(t;\boldsymbol\eta)-\Psi(t;\mathbf0) =\sum_{\boldsymbol\alpha\ne\mathbf0}
  \frac{\boldsymbol\eta^{\boldsymbol\alpha}}{\boldsymbol\alpha!} \Psi_{\boldsymbol\alpha}(t),
  \label{app:eq:Psi_multivariate_Taylor}
\end{equation}
where the partial derivatives are defined in Eq.~\eqref{eq:radial_derivative_blocks}.
Exponentiating this series gives
\begin{equation*}
  \prod_{\boldsymbol\alpha\ne\mathbf0} \sum_{n_{\boldsymbol\alpha}\geq0}
  \frac{[\boldsymbol\eta^{\boldsymbol\alpha}
    \Psi_{\boldsymbol\alpha}(t)]^{n_{\boldsymbol\alpha}}}
  {n_{\boldsymbol\alpha}!\,(\boldsymbol\alpha!)^{n_{\boldsymbol\alpha}}}.
\end{equation*}
Each term contains the monomial
\(\boldsymbol\eta^{\sum_{\boldsymbol\alpha\ne\mathbf0} n_{\boldsymbol\alpha}\boldsymbol\alpha}\).
Selecting \(\boldsymbol\eta^{\mathbf p}\) therefore imposes the configuration constraint in
Eq.~\eqref{eq:radial_configuration_condition} and gives the factors in
Eq.~\eqref{eq:radial_configuration_factors}.
The factors \(\boldsymbol\alpha!\) and \(n_{\boldsymbol\alpha}!\) originate from the Taylor
and exponential expansions, respectively.
For fixed \(\mathbf p\), only finitely many configurations satisfy the constraint, yielding
Eq.~\eqref{eq:finite_rho_expansion}.

\raggedbottom
\begingroup
\section{Numerical coefficient extraction by Cauchy sampling}
\label{app:cauchy-extraction}
After analytic transverse integration and OAM extraction, Cauchy
sampling~\cite{Bornemann2011Cauchy} avoids the costly explicit radial expansion at high
orders.
For fixed OAM indices, define the longitudinally integrated generating function and its
coefficients
\begin{equation}
  \begin{aligned}
    \mathcal H(\boldsymbol\eta) ={}&\int_{-1}^{1}dt\,\widetilde\chi(t)\mathcal G(t)
    \exp[\Psi(t;\boldsymbol\eta)-\Psi(t;\mathbf0)], \\
    h_{\mathbf p} ={}&[\eta_{\mathrm p}^{p_{\mathrm p}}\eta_{\mathrm s}^{p_{\mathrm s}}
    \eta_{\mathrm i}^{p_{\mathrm i}}]\mathcal H(\boldsymbol\eta).
  \end{aligned}
  \label{app:eq:integrated-radial-generator}
\end{equation}
Multiplying \(h_{\mathbf p}\) by the OAM selection factor and the prefactor in
Appendix~\ref{app:overall-prefactor} gives the amplitude in
Eq.~\eqref{eq:final_normalized_amplitude}.
We evaluate the rational generating kernel directly with Gauss--Legendre longitudinal
quadrature, using complex double precision for the sampled benchmarks.

For a common radius \(0<r_{\mathrm c}<1\) within the domain of analyticity and
\(N>\max(p_{\mathrm p},p_{\mathrm s},p_{\mathrm i})\), sample each variable at
\(\eta_j=r_{\mathrm c}e^{i\theta_m}\), where \(\theta_m=2\pi m/N\) and \(m=0,\ldots,N-1\).
The periodic trapezoidal rule gives the discrete Fourier form of Cauchy's coefficient
formula,
\begin{equation}
  \makebox[0.8\columnwidth][c]{\resizebox{0.90\columnwidth}{!}{$\displaystyle
      \begin{aligned}
        &h_{\mathbf p}^{(N,r_{\mathrm c})}
        =\frac{1}{N^3r_{\mathrm c}^{p_{\mathrm p}+p_{\mathrm s}+p_{\mathrm i}}}
        \sum_{m_{\mathrm p},m_{\mathrm s},m_{\mathrm i}=0}^{N-1} \\
        &\quad\times\mathcal H\!\left(r_{\mathrm c}e^{i\theta_{m_{\mathrm p}}},
        r_{\mathrm c}e^{i\theta_{m_{\mathrm s}}}, r_{\mathrm c}e^{i\theta_{m_{\mathrm i}}}\right)
        \times e^{-i(p_{\mathrm p}\theta_{m_{\mathrm p}} +p_{\mathrm s}\theta_{m_{\mathrm s}}
          +p_{\mathrm i}\theta_{m_{\mathrm i}})}.
      \end{aligned}
  $}}
  \label{app:eq:cauchy-coefficient-sampling}
\end{equation}
For \(p_{\mathrm p}=0\), setting \(\eta_{\mathrm p}=0\) selects the pump coefficient exactly
and reduces the sampling to two variables.
The same samples yield a matrix of signal--idler radial coefficients.

For exact longitudinal integration, the Fourier aliasing is
\begin{equation}
  h_{\mathbf p}^{(N,r_{\mathrm c})}-h_{\mathbf p}
  =\sum_{\boldsymbol v\in\mathbb N_0^3\setminus\{\mathbf0\}} h_{\mathbf p+N\boldsymbol v}\,
  r_{\mathrm c}^{N(v_{\mathrm p}+v_{\mathrm s}+v_{\mathrm i})}.
  \label{app:eq:cauchy-aliasing}
\end{equation}
The numerical radius \(r_{\mathrm c}\) balances suppression of aliased coefficients against
round-off amplification by \(r_{\mathrm c}^{-(p_{\mathrm p}+p_{\mathrm s}+p_{\mathrm i})}\).
We check convergence by increasing angular samples and longitudinal quadrature nodes and
varying \(r_{\mathrm c}\), with converged coefficients independent of the radius.
Benchmark results are given in Appendix~\ref{app:runtime-benchmark}.
\endgroup

\section{Overall prefactor}
\label{app:overall-prefactor}
With the LG normalization factors defined after Eq.~\eqref{eq:LG_from_generating_function},
the product for the pump and conjugated signal and idler modes is
\begin{equation}
  \makebox[0.8\columnwidth][c]{\resizebox{0.90\columnwidth}{!}{$\displaystyle
      \begin{aligned}
        &N^{\ell_{\mathrm p}}_{p_{\mathrm p}}(w_{\mathrm p})
        \left[N^{\ell_{\mathrm s}}_{p_{\mathrm s}}(w_{\mathrm s})\right]^{*}
        \left[N^{\ell_{\mathrm i}}_{p_{\mathrm i}}(w_{\mathrm i})\right]^{*} =(-1)^c
        \frac{w_{\mathrm p}^3\,2^{a+b+c}} {\gamma_{\mathrm s}\gamma_{\mathrm i}(2\pi)^{3/2}} \\
        &\qquad\times\left[ \frac{p_{\mathrm p}!p_{\mathrm s}!p_{\mathrm i}!}
        {(p_{\mathrm p}+|\ell_{\mathrm p}|)!(p_{\mathrm s}+|\ell_{\mathrm s}|)!(p_{\mathrm i}+|\ell_{\mathrm i}|)!}
        \right]^{1/2}.
      \end{aligned}
  $}}
  \label{app:eq:LG_norm_product}
\end{equation}
Combining Eq.~\eqref{app:eq:LG_norm_product}, the global normalization \(\mathcal N\), the
transverse Gaussian integral in Eq.~\eqref{eq:transverse_gaussian_result}, and the OAM
derivative factors gives the integration-independent prefactor in
Eq.~\eqref{eq:final_normalized_amplitude},
\begin{equation}
  \makebox[0.8\columnwidth][c]{\resizebox{0.90\columnwidth}{!}{$\displaystyle
      \begin{aligned}
        &\mathcal A_{p_{\mathrm p},p_{\mathrm s},p_{\mathrm i}}^{\ell_{\mathrm p},\ell_{\mathrm s},\ell_{\mathrm i}}
        =\mathcal N\frac{\sqrt{2\pi}L}{w_{\mathrm p}}
        \frac{|\ell_{\mathrm p}|!\,|\ell_{\mathrm s}|!\,|\ell_{\mathrm i}|!}{a!\,b!\,c!}
        \times\frac{ 2^{a+b+c+1}\gamma_{\mathrm s}^{a+c+1}\gamma_{\mathrm i}^{b+c+1}
        }{(\gamma_{\mathrm s}^2+\gamma_{\mathrm i}^2+1)^{a+b+c+1}} \\
        &\quad\times\left[ \frac{p_{\mathrm p}!p_{\mathrm s}!p_{\mathrm i}!}
        {(p_{\mathrm p}+|\ell_{\mathrm p}|)!(p_{\mathrm s}+|\ell_{\mathrm s}|)!(p_{\mathrm i}+|\ell_{\mathrm i}|)!}
        \right]^{1/2}.
      \end{aligned}
  $}}
  \label{app:eq:prefactor}
\end{equation}

\section{Direct radial trace}
\label{app:direct-radial-trace}
For a fixed pump mode and OAM indices
\((\ell_{\mathrm p},\ell_{\mathrm s},\ell_{\mathrm i})\), extracting the pump radial and
three OAM coefficients from Eq.~\eqref{eq:generating_amplitude} gives
\begin{equation}
  \begin{gathered}
    \mathscr{Z}_{p_{\mathrm p}}^{\ell_{\mathrm p},\ell_{\mathrm s},\ell_{\mathrm i}}(\eta_{\mathrm s},\eta_{\mathrm i})
    =\sum_{p_{\mathrm s},p_{\mathrm i}\geq0}
    \widetilde C_{p_{\mathrm s},p_{\mathrm i}}^{\ell_{\mathrm p},\ell_{\mathrm s},\ell_{\mathrm i}}\eta_{\mathrm s}^{p_{\mathrm s}}\eta_{\mathrm i}^{p_{\mathrm i}}, \\
    C_{p_{\mathrm p},p_{\mathrm s},p_{\mathrm i}}^{\ell_{\mathrm p},\ell_{\mathrm s},\ell_{\mathrm i}}
    =\left[ \frac{p_{\mathrm s}!p_{\mathrm i}!}
    {(p_{\mathrm s}+|\ell_{\mathrm s}|)!(p_{\mathrm i}+|\ell_{\mathrm i}|)!} \right]^{1/2}
    \widetilde C_{p_{\mathrm s},p_{\mathrm i}}^{\ell_{\mathrm p},\ell_{\mathrm s},\ell_{\mathrm i}},
  \end{gathered}
  \label{app:eq:radial-generating-amplitude}
\end{equation}
where \(\mathscr{Z}_{p_{\mathrm p}}^{\ell_{\mathrm p},\ell_{\mathrm s},\ell_{\mathrm i}}\)
is the generating amplitude for the signal and idler radial indices, and
\(\widetilde C_{p_{\mathrm s},p_{\mathrm i}}^{\ell_{\mathrm p},\ell_{\mathrm s},\ell_{\mathrm i}}\)
are its Taylor coefficients, including all factors independent of \(p_{\mathrm s}\) and
\(p_{\mathrm i}\).
The physical amplitudes \(C\) follow by applying the radial normalization factors in the
second line.
Using the weight measure \(dW_m(u)\), the radial normalization factors satisfy
\begin{equation*}
  \resizebox{0.99\columnwidth}{!}{$\displaystyle
    \frac{p!}{(p+m)!} =\int_0^1 u^p\,dW_m(u), \quad dW_m(u)=\frac{(1-u)^{m-1}}{(m-1)!}\,du
    \quad(m>0).
  $}
\end{equation*}
For \(m=0\), \(dW_0(u)=\delta(1-u)du\).
The two-dimensional Parseval identity applied to
Eq.~\eqref{app:eq:radial-generating-amplitude} gives, for \(0<\tau<1\),
\begin{equation*}
  \resizebox{0.99\columnwidth}{!}{$\displaystyle
    \begin{aligned}
      &\mathcal T_{\tau}^{\ell_{\mathrm s},\ell_{\mathrm i}}\!(p_{\mathrm p},\ell_{\mathrm p})
      =\int\!dW_{|\ell_{\mathrm s}|}(u_{\mathrm s}) \int\!dW_{|\ell_{\mathrm i}|}(u_{\mathrm i})
      \int_0^{2\pi}\frac{d\theta_{\mathrm s}}{2\pi} \int_0^{2\pi}\frac{d\theta_{\mathrm i}}{2\pi} \\
      &\quad\times\left|
      \mathscr{Z}_{p_{\mathrm p}}^{\ell_{\mathrm p},\ell_{\mathrm s},\ell_{\mathrm i}}
      \left(\tau\sqrt{u_{\mathrm s}}e^{i\theta_{\mathrm s}},
      \tau\sqrt{u_{\mathrm i}}e^{i\theta_{\mathrm i}}\right) \right|^2 \\
      &=\sum_{p_{\mathrm s},p_{\mathrm i}\geq0}
      \left|C_{p_{\mathrm p},p_{\mathrm s},p_{\mathrm i}}^{\ell_{\mathrm p},\ell_{\mathrm s},\ell_{\mathrm i}}\right|^2
      \tau^{2(p_{\mathrm s}+p_{\mathrm i})}.
    \end{aligned}
  $}
\end{equation*}
The radially insensitive OAM spectrum in Eq.~\eqref{eq:radially_insensitive_oam_spectrum}
follows from the Abel limit
$P_{\mathrm{tr}}^{\ell_{\mathrm s},\ell_{\mathrm i}}(p_{\mathrm p},\ell_{\mathrm p}) =\lim_{\tau\to1^-}\mathcal T_{\tau}^{\ell_{\mathrm s},\ell_{\mathrm i}}(p_{\mathrm p},\ell_{\mathrm p})$,
without individual radial coefficient extraction or a radial cutoff.

For the main-text spectra, we linearly extrapolate $\mathcal T_{\tau}$ from $\tau=0.990$ and
$0.995$ in $\sqrt{1-\tau}$, using 96 uniformly spaced samples per angular integral,
224-point Gauss--Legendre quadrature for $t$, and 14-point Gauss--Jacobi quadrature per
nontrivial beta integral.

Refinement to 192 angular samples, 448 longitudinal nodes, and 24 beta nodes changes the
results at
$(\gamma,\phi,\ell_{\mathrm p},\ell_{\mathrm s},\ell_{\mathrm i})=(0.5,0.10,0,0,0)$ and
$(5.0,1.50,3,0,3)$ by $0.80\%$ and $1.59\%$, respectively.
At $\tau=0.65$, explicit radial sums through $p_{\mathrm s},p_{\mathrm i}\leq16$ agree with
the direct trace within $2.6\times10^{-13}$ and $1.1\times10^{-10}$ for zero-OAM and
same-sign OAM tests, respectively.
In a nondegenerate high-order test, exchanging all signal and idler mode labels, waists, and
wave numbers changes the amplitude by a relative $8.4\times10^{-58}$.

\section{Numerical validation of the dimensionless reduction}
\label{app:numerical-validation}
\raggedbottom

\subsection{Validation of scale invariance}
\label{app:universal-collapse}

We test the scale invariance implied by the dimensionless parameters in
Eq.~\eqref{eq:geometric_controls} using
\begin{equation}
  \makebox[0.8\columnwidth][c]{\resizebox{0.90\columnwidth}{!}{$\displaystyle
      L\longrightarrow sL, \qquad
      w_j\longrightarrow \sqrt{s}\,w_j,\quad s\in\{0.50,0.75,1.00,1.50\}.
  $}}
  \label{app:eq:collapse-rescaling}
\end{equation}
This rescaling preserves the waist ratios, focusing parameters $L/(k_jw_j^2)$, and
wave-number ratio $\nu$, so the normalized LG spectrum is unchanged.
We use \texttt{SPDCAmplitude.jl}, based on Ref.~\cite{PhysRevA.106.063711}, with the
dimensional inputs in Table~\ref{tab:collapse-parameters}.
The Sevilla--10 and Sevilla--20 inputs follow Ref.~\cite{SevillaGutierrez2024}, and
Kovlakov, Valencia, and Villar follow Refs.~\cite{Kovlakov2018}, \cite{Valencia2021}, and
\cite{Villar2019}, respectively.
These representative inputs do not constitute complete models of the experimental sources.
For all three fields, $k_j=2\pi n_j/\lambda_j$, with $n_j=1.8444$ for Sevilla and Kovlakov,
$1.816$ for Valencia, and $1.66$ for Villar.
\par\smallskip\noindent
\begin{minipage}{\columnwidth}
  \captionsetup{skip=4pt}
  \captionof{table}{Dimensional inputs for the scaling test.
    The Valencia collection waist follows from $\gamma=5.26$.
    Villar uses the circular-pump configuration.}
  \label{tab:collapse-parameters}
  \centering
  \footnotesize
  \setlength{\tabcolsep}{2.5pt}
  \begin{tabular}{@{}lrrrrrr@{}}
    \toprule
    Set & \shortstack{\(L\)\\(mm)} & \shortstack{\(w_{\mathrm p}\)\\(\(\mu\)m)}
    & \shortstack{\(w_{\mathrm s}=w_{\mathrm i}\)\\(\(\mu\)m)}
    & \shortstack{\(\lambda_{\mathrm p}\)\\(nm)}
     & \shortstack{\(\lambda_{\mathrm s}/\lambda_{\mathrm i}\)\\(nm)} & \(\gamma\) \\
    \midrule
    Sevilla--10 & 10 & 141 & 41.0 & 405 & 810/810   & 3.439 \\
    Sevilla--20 & 20 & 60  & 30.0 & 405 & 810/810   & 2.000 \\
    Kovlakov    & 15 & 25  & 33.0 & 405 & 810/810   & 0.758 \\
    Valencia    & 5  & 450 & 85.6 & 775 & 1550/1550 & 5.260 \\
    Villar      & 5  & 93  & 38.0 & 405 & 776/847   & 2.447 \\
    \bottomrule
  \end{tabular}
\end{minipage}
\par\smallskip

\begin{figure*}[t]
  \centering
  \includegraphics[width=0.90\textwidth]{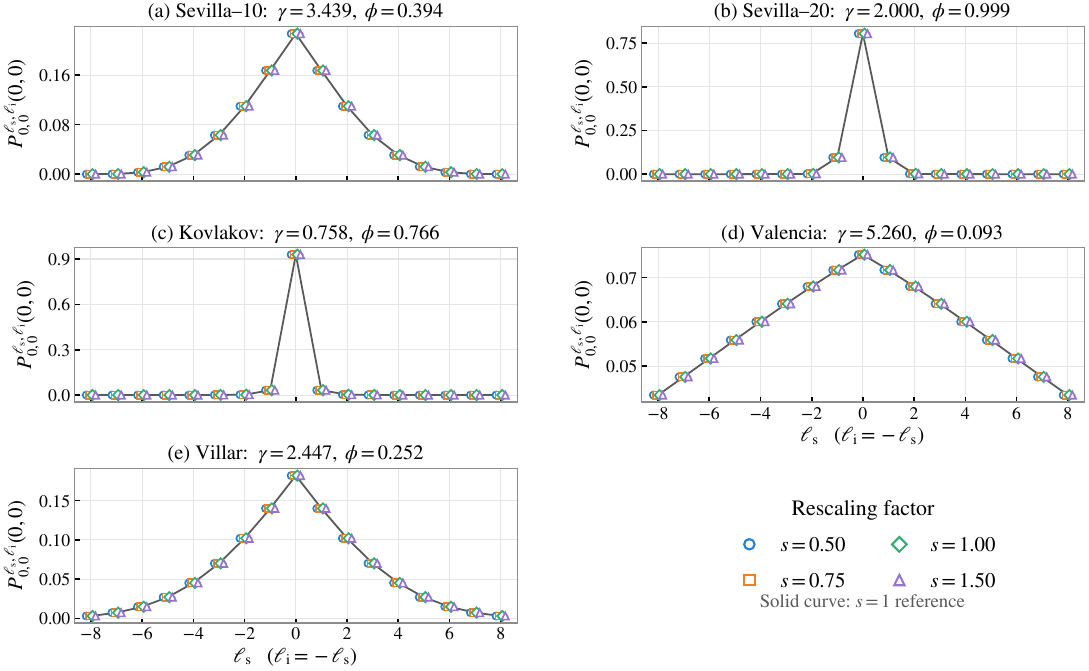}
  \caption{Scale
    invariance under Eq.~\eqref{app:eq:collapse-rescaling}.
    Each $\ell_{\mathrm p}=0$ spectrum is normalized over $-8\leq\ell_{\mathrm s}\leq8$, with
    $\ell_{\mathrm i}=-\ell_{\mathrm s}$.
    The solid curve denotes $s=1$.
    Open markers denote all four rescalings, with horizontal offsets for visibility.
    Probability differences are listed in Table~\ref{tab:collapse-errors}.}
  \label{fig:universal-collapse-validation}
\end{figure*}

The rescaled spectra overlap in Fig.~\ref{fig:universal-collapse-validation}.
Table~\ref{tab:collapse-errors} compares the $p_{\mathrm s}=p_{\mathrm i}=0$ OAM spectra for
$\ell_{\mathrm p}=0,2$ and the radial probability matrices for
$0\leq p_{\mathrm s},p_{\mathrm i}\leq3$.
The maximum absolute probability difference is $3.76\times10^{-12}$.
\par\smallskip\noindent
\begin{minipage}{\columnwidth}
  \captionsetup{skip=4pt}
  \captionof{table}{Maximum absolute probability differences among the four dimensional
    rescalings.}
  \label{tab:collapse-errors}
  \centering
  \footnotesize
  \setlength{\tabcolsep}{3pt}
  \begin{tabular}{@{}lccc@{}}
    \toprule
    Set         & \(\ell_{\mathrm p}=0\)     & \(\ell_{\mathrm p}=2\)     & radial \(4\times4\)        \\
    \midrule
    Sevilla--10 & \(9.36\!\times\!10^{-15}\) & \(1.10\!\times\!10^{-12}\) & \(2.04\!\times\!10^{-13}\) \\
    Sevilla--20 & \(1.60\!\times\!10^{-15}\) & \(2.27\!\times\!10^{-14}\) & \(2.82\!\times\!10^{-14}\) \\
    Kovlakov    & \(4.01\!\times\!10^{-16}\) & \(1.13\!\times\!10^{-15}\) & \(1.22\!\times\!10^{-14}\) \\
    Valencia    & \(1.76\!\times\!10^{-14}\) & \(3.76\!\times\!10^{-12}\) & \(1.17\!\times\!10^{-13}\) \\
    Villar      & \(3.98\!\times\!10^{-15}\) & \(1.63\!\times\!10^{-13}\) & \(8.08\!\times\!10^{-14}\) \\
    \bottomrule
  \end{tabular}
\end{minipage}
\par\smallskip

\par\smallskip\noindent
\begin{minipage}{\columnwidth}
  \captionsetup{skip=4pt}
  \captionof{table}{Runtime comparison. ``Failed'' and dashes follow the definitions in
    Table~\ref{tab:main-runtime-comparison}.}
  \label{tab:runtime-benchmark}
  \centering
  \small
  \setlength{\tabcolsep}{3pt}
  \begin{tabular*}{\linewidth}{@{\extracolsep{\fill}}lrrr@{}}
    \toprule
    Order          & Previous (s) & This work (s)         & Speedup        \\
    \midrule
    \multicolumn{4}{@{}l}{Only radial: Sevilla--10 (degenerate)} \\
    \midrule
    \(p=10\)       & 0.114        & 0.00571               & \(19.9\times\) \\
    \(p=11\)       & 0.116        & 0.00561               & \(20.7\times\) \\
    \(p=15\)       & failed       & 0.00561               & --             \\
    \(p=20\)       & failed       & 0.0168                & --             \\
    \(p=30\)       & failed       & 0.0970                & --             \\
    \(p=40\)       & failed       & 0.0973                & --             \\
    \midrule
    \multicolumn{4}{@{}l}{Only OAM: Sevilla--10 (degenerate)} \\
    \midrule
    \(|\ell|=30\)  & 0.167        & \(5.35\times10^{-4}\) & \(311\times\)  \\
    \(|\ell|=35\)  & failed       & \(7.04\times10^{-4}\) & --             \\
    \(|\ell|=40\)  & failed       & \(9.58\times10^{-4}\) & --             \\
    \(|\ell|=120\) & failed       & 0.00601               & --             \\
    \bottomrule
  \end{tabular*}
\end{minipage}
\par\smallskip

\subsection{Runtime and numerical-stability comparison}
\label{app:runtime-benchmark}
Tables~\ref{tab:main-runtime-comparison} and~\ref{tab:runtime-benchmark} compare the
runtimes of \texttt{SPDCAmplitude.jl} (``Previous'') and \texttt{SPDCForge.jl} (``This
work'') with the same amplitude normalization.
Errors are relative amplitude differences.
The radial and OAM family is defined by
\begin{equation}
  \makebox[0.8\columnwidth][c]{\resizebox{0.90\columnwidth}{!}{$\displaystyle
      (p_{\mathrm p},p_{\mathrm s},p_{\mathrm i},\ell_{\mathrm p},\ell_{\mathrm s},\ell_{\mathrm i}) =(n-1,n,n,2n,3n+1,-n-1).
  $}}
  \label{app:eq:mixed-benchmark-family}
\end{equation}
The degenerate and nondegenerate benchmarks use the Sevilla--10 and Villar parameters in
Table~\ref{tab:collapse-parameters}, respectively.
The radial family uses $(p_{\mathrm p},p_{\mathrm s},p_{\mathrm i})=(0,p,p)$ with zero OAM.

For Sevilla--10, the relative amplitude errors are \mbox{$3.33\times10^{-6}$} at $n=2$,
$6.05\times10^{-5}$ and $1.42\times10^{-3}$ at $p=10$ and $11$, respectively, and
$6.12\times10^{-6}$ at $|\ell|=30$.
For Villar, the error is $1.61\times10^{-7}$ at $n=2$.

\flushbottom
\bibliographystyle{apsrev4-2}
\bibliography{reference}

\end{document}